\documentclass{aa}  
\usepackage{subcaption}
\usepackage{color}
\usepackage{natbib,twoopt}
\usepackage[hyphenbreaks]{breakurl}
\usepackage[breaklinks]{hyperref}      
\bibpunct{(}{)}{;}{a}{}{,}             
\definecolor{cobalt}{rgb}{0.06, 0.2, 0.65}
\hypersetup{
  colorlinks,
  citecolor=cobalt,
  linkcolor=[rgb]{0.8, 0.2, 1.0},
  urlcolor=cobalt,
}
\makeatletter
  \newcommandtwoopt{\citeads}[3][][]{\href{http://adsabs.harvard.edu/abs/#3}%
    {\def\hyper@linkstart##1##2{}%
     \let\hyper@linkend\@empty\citealp[#1][#2]{#3}}}
  \newcommandtwoopt{\citepads}[3][][]{\href{http://adsabs.harvard.edu/abs/#3}%
    {\def\hyper@linkstart##1##2{}%
     \let\hyper@linkend\@empty\citep[#1][#2]{#3}}}
  \newcommandtwoopt{\citealtads}[3][][]{\href{http://adsabs.harvard.edu/abs/#3}%
    {\def\hyper@linkstart##1##2{}%
     \let\hyper@linkend\@empty\citealt[#1][#2]{#3}}}
  \newcommandtwoopt{\citeyearads}[3][][]%
    {\href{http://adsabs.harvard.edu/abs/#3}
    {\def\hyper@linkstart##1##2{}%
     \let\hyper@linkend\@empty\citeyear[#1][#2]{#3}}}
\makeatother

\usepackage{graphicx}
\usepackage{txfonts}
\begin{document} 

   \title{The imprints of galaxy cluster internal dynamics on the Sunyaev-Zeldovich effect}

   \author{Óscar Monllor-Berbegal\inst{1}\thanks{oscar.monllor@uv.es},
          David Vallés-Pérez\inst{1},
          Susana Planelles\inst{1,2},
          Vicent Quilis\inst{1,2}
          }

   \institute{Departament d’Astronomia i Astrofísica, Universitat de València,
              46100 Burjassot (València), Spain
         \and
            Observatori Astron\`omic, Universitat de Val\`encia, E-46980 Paterna (València), Spain}

   \date{Received X XX, XXXX; accepted X XX, XXXX}

  \abstract
  {Forthcoming measurements of the Sunyaev-Zeldovich (SZ) effect in galaxy clusters will dramatically improve our understanding of the main intra-cluster medium (ICM) properties and how they depend on the particular thermal and dynamical state of the associated clusters.}
   {Using a sample of simulated galaxy clusters, whose dynamical history can be well known and described, we assess the impact of the ICM internal dynamics on both the thermal and kinetic SZ effects (tSZ and kSZ, respectively).}
   {We produce synthetic maps of the SZ effect, both thermal and kinetic, for the simulated clusters obtained in a cosmological simulation produced by a cosmological Adaptive Mesh Refinement (AMR) code. For each galaxy cluster in the sample, its dynamical state is estimated by using a combination of well-established indicators. 
   We use the correlations between SZ maps and cluster dynamical state, to look for the imprints of the evolutionary events, mainly mergers, on the SZ signals.}
   {While the tSZ effect only shows dependency on dynamical state in its radial distribution, the kinetic effect shows a remarkable correlation with this property: unrelaxed clusters present a higher radial profile and an overall stronger signal at all masses and radii. 
   The reason for this correlation is the fuzziness of the ICM produced by recent merging episodes. Furthermore, the kSZ signal is correlated with rotation for relaxed clusters, while for the disturbed systems the effect is dominated by other motions such as bulk flows, turbulence, etc. The kSZ effect shows a dipolar pattern when averaging over cluster dynamical classes, especially for the relaxed population. This feature can be exploited to stack multiple kSZ maps in order to recover a stronger dipole signal that would be correlated with the global rotation properties of the sample.}
   {The SZ effect can be used as a tool to estimate the dynamical state of galaxy clusters, especially to segregate those clusters with a quiescent evolution from those with a rich record of recent merger events. Our results suggest that the forthcoming observational data measuring the SZ signal in clusters could be used as a complementary strategy to classify the evolutionary history of galaxy clusters.}


   \titlerunning{Galaxy cluster internal dynamics and the Sunyaev-Zeldovich effect}
   \authorrunning{Monllor-Berbegal, Vallés-Pérez, Planelles, Quilis}

  \maketitle
%

\section{Introduction}
\label{s:intro}

Galaxy clusters (GCs) are collections of 50 to a few thousand galaxies held together by gravitational forces. These clusters have typical masses around $10^{14} \; \mathrm{M_{\odot}}$ and radius of \mbox{$R \gtrsim 1 \; \mathrm{Mpc}$}. The majority of the mass in GCs, up to \mbox{84-90\%}, is in the form of dark matter (DM), while only 10-16\% is in the form of baryonic matter, primarily hot gas known as the intra-cluster medium (ICM). Only 1-5\% of the mass is composed of stars \citep{cimatti2019introduction}.

The ICM is an almost fully ionised plasma at temperatures around $T_\mathrm{ICM} \sim 10^{7-8} \, \mathrm{K}$. This plasma extends throughout the entire cluster volume and emits X-rays predominantly through thermal bremsstrahlung \citep{sarazin1988book, birkinshaw1999sunyaev}. Electrons in the ICM are not only scattered by ions (bremsstrahlung), but can themselves scatter photons. Hence, apart from being extremely luminous X-ray sources, GCs also leave an imprint on the cosmic microwave background (CMB; \citealt{penzias1965measurement}) through the Sunyaev-Zeldovich effect (SZ; \citealt{zeldovich1969interaction}), arising from the scattering of CMB photons off ICM electrons through the inverse Compton effect \citep[see][for reviews]{rephaeli1995comptonization, birkinshaw1999sunyaev, carlstrom2002cosmology,mroczkowski2019astrophysics}.

The SZ effect can be separated in two main components, the {thermal (tSZ)} and the {kinetic (kSZ)} effects. The thermal signal is produced as CMB photons travel through the ICM of a GC and get scattered, in any direction, by the random thermal motions of the ICM electrons and, since the ICM is very hot, it shifts the CMB spectrum towards higher frequencies. The change in the black-body CMB temperature ($T$) due to the tSZ effect (first calculated by \citealt{zeldovich1969interaction} using the \citealt{kompaneets1957establishment} equation) is given by \cite{mroczkowski2019astrophysics}:
\begin{equation}
\label{tSZ}
    \frac{\Delta T_\mathrm{tSZ}}{T} = f(\nu, T) \, y_\mathrm{tSZ} ,\;\;\;\;\; \text{with}\;\;\;\;\; y_\mathrm{tSZ} = \frac{\sigma_T}{m_e c^2} \int \mathrm{d}s \; {k_B T_e n_e} \;\;\;\;   
\end{equation}
where $n_e$ is the electron number density, $T_e$ the electron gas temperature, $m_e$ the electron mass, $\sigma_T$ the Thomson scattering cross-section, $k_B$ the Boltzmann constant, and $c$ the speed of light. On the other hand, $f(\nu, T)$ is a function of frequency and \mbox{CMB temperature}\footnote{The expression for $f$ can be the original, non-relativistic, diffusive solution provided by \cite{zeldovich1969interaction} or the one including the  more complicated relativistic corrections by \cite{rephaeli1995comptonization}.}, $y_\mathrm{tSZ}$ is the so called Compton $y$-parameter, that encapsulates the GC temperature and density distributions and $\mathrm{d}s$ is the physical distance travelled by the photon along the line-of-sight. We remark that considering the electron gas as an ideal gas (which due to the temperature and density regime is accurate enough), the equation of state is $P_e = k_B T_e n_e$ and, therefore, $y_\mathrm{tSZ}$ measures the \mbox{electron pressure ($P_e$)} integrated along the line-of-sight.

On the other hand, the kinetic effect arises from the fact that, due to the bulk motion of the GC and the internal movements of the ICM (rotation, turbulence, etc.),  the scattering medium is moving with respect to the CMB reference frame (Hubble flow). The change in the CMB temperature due to the kSZ effect alone is given by \citep{sunyaev1980velocity, mroczkowski2019astrophysics}:
\begin{equation}
\label{kSZ}
    \frac{\Delta T_\mathrm{kSZ}}{T} = - y_\mathrm{kSZ}, \;\;\;\;\; \text{with}\;\;\;\;\; y_\mathrm{kSZ} = \sigma_T \int \mathrm{d}s \frac{v_\mathrm{los}}{c} n_e   \;\;\;\;   
\end{equation}
where $v_\mathrm{los}$ is the velocity along the line-of-sight and $y_\mathrm{kSZ}$ is the Compton $y$-parameter for the kSZ effect. The sign criteria is such that if the gas is moving away from the observer ($v_\mathrm{los} > 0$), it decreases the CMB temperature. In this case, $y_\mathrm{kSZ}$ encapsulates the kinematic structure of the cluster and its peculiar motion. Contrarily  to the tSZ signal, the kinetic effect does not depend on frequency and, therefore, it cannot modify the CMB black-body spectrum. In this sense, the kSZ effect can be understood as a Doppler boost of CMB photons \citep[e.g.][]{baxter2019constraining}. In fact, as \cite{rephaeli1995comptonization} pointed out, Eq.~\eqref{kSZ} can be obtained from a simple relativistic transformation. 

The kSZ signal can be further split in different contributions. The one that contributes the most is the bulk or monopole component \citep[e.g.][]{hand2012evidence, chen2022detection}, which is due to the movement of the cluster as a whole with respect to the CMB. Nevertheless, if we subtract this component, we get more contributions due to the ICM internal motions such as rotation, which produces a dipole-like signal \citep[e.g.][]{cooray2002kinetic, chluba2002kinetic}, or random motions due to turbulence which fill most of the ICM volume (e.g. see \citealt{vazza2017turbulence, valles2021troubled, valles2021unravelling}).  While the monopole can be used to measure GC peculiar velocities, the other contributions are useful to understand the internal dynamics of the gas in GCs. The kinetic SZ signal remaining from the subtraction of the monopole has been studied especially for an ideal rotation alignment, that is, when the angular momentum is perpendicular to the line-of-sight. In this particular case, the dipole signal due to rotation maximises and the effect is known as the rotational kSZ signal \citep[rkSZ;][]{cooray2002kinetic, chluba2002kinetic}. However, in general, the rotation axes will be randomly aligned, and hence, the dipole signal due to rotation will not necessarily be the dominant contribution after the monopole. Other ICM internal movements, such as turbulence or outflows, could contribute significantly.

Whereas the frequency-dependent tSZ effect is the dominant component and is relatively easy to observe, the kSZ effect, which generates a fainter frequency-independent signal, can be easily confused with primary CMB anisotropies {\citep[e.g.][]{Colafrancesco_2007}}. Nevertheless, on small angular scales, where these anisotropies have minimal influence, the kSZ effect enables the measurement of gas peculiar velocities, particularly velocity gradients within the ICM (see \citealt{dupke2002prospects} for a discussion on these measurements). Hence, in combination with X-ray emission and the tSZ signal, the kSZ effect can help disentangling the complex ICM velocity structure produced by AGN feedback, cosmological structure formation or turbulent motions \citep[see, e.g.][]{battaglia2017future, simionescu2019constraining}. Moreover, since the kSZ signal is temperature independent and linearly proportional to electron density, it is especially convenient to explore low density and temperature regions such as the outskirts of galaxy clusters and groups of galaxies, thus having the potential of shedding light on the \textit{missing baryon problem} (e.g. see \citealt{hernandez2015evidence} for observational evidence or \citealt{planelles2018multiwavelength} for a computational approach). From a different point of view, the SZ effect measurements can be used to disentangle the nature of other components of the Universe such as dark matter or dark energy, as well as a tool to test the feasibility of different cosmological models (e.g. non-flat Universe) or other gravity theories \citep[see, e.g.][]{alfano2023breaking}.

Currently, thanks to measurements carried out by observational facilities such as \textsc{Planck} \citep{ade2014planck}, South Pole Telescope (\textsc{SPT}; \citealt{reichardt2013galaxy}), Atacama Cosmology Telescope  (\textsc{ACT}; \citealt{hasselfield2013atacama}) or Combined Array for Research in Millimeter-wave Astronomy (\textsc{CARMA} ; \citealt{plagge2013carma}), the number of clusters studied using the SZ signal has increased significantly. Moreover, in the upcoming decades future X-ray missions, such as \textsc{Xrism}\footnote{\url{https://xrism.isas.jaxa.jp/en/}} and \textsc{Athena}\footnote{\url{https://www.the-athena-x-ray-observatory.eu/en}}, are expected to provide high precision measurements of ICM motions. 
In a similar way, future mm/sub-mm facilities (see, for instance, \textsc{AtLAST}\footnote{\url{https://www.atlast.uio.no/}} or \textsc{SKA}\footnote{\url{https://www.skao.int/}}) are expected to have enough resolution to be able to measure the ICM velocity field by means of the kSZ effect.
This synergy between future X-ray and mm/sub-mm observations will provide us unprecedented insight on the ICM thermodynamic properties.


Many successful attempts to measure bulk motions with the kSZ effect have been carried out so far \citep[e.g.][]{hand2012evidence, soergel2016detection, hill2016kinematic, chen2022detection}. Concerning the internal movements contribution to the kSZ effect, an early attempt to produce a spatially resolved map of the kSZ signal in a galaxy cluster was presented by \cite{sayers2013measurement}. An improved measurement was provided by
\cite{adam2017mapping}, who detected a dipolar structure in the kSZ signal of the merging cluster MACS J0717.5+3745. The signal was consistent with two sub-clusters, moving with velocities of several thousand $\mathrm{km}\,\mathrm{s}^{-1}$, one moving away and the other approaching the observer. In \cite{matilla2020probing}, the authors study the viability of statistically detecting a rotational kSZ signal through the combination of CMB data across numerous galaxies, where the spin orientation can be spectroscopically estimated. 

Other recent observational efforts to further study gas thermodynamics with both thermal and kinetic SZ are \cite{schaan2016evidence, schaan2021atacama}, where the authors use both effects to study the baryon density and pressure profiles of galactic haloes, finding that they are shallower than their DM counterparts. In a similar direction, \cite{ACT_2021} constrain the amount of non-thermal pressure inside the virial radius of GCs and \cite{Mallaby-Kay_2023} present a novel hybrid estimator to measure kSZ, putting constraints on the thermodynamic properties of galaxy haloes. Furthermore, regarding the cluster dynamical state and SZ connection, \cite{adam2023xxl} confirm that the observed disturbed clusters have less concentrated and shallower electron pressure profiles compared to what would be expected for a relaxed cluster.


Early works using numerical simulations, such as \cite{Springel_2001}, already put constraints on the mean Comptonization due to tSZ. More recent works \citep[e.g.][]{planelles2017pressure, planelles2018multiwavelength, lokken2023boundless} have further expanded our knowledge studying the tSZ effect on clusters and their surroundings producing mock $y$ maps and scaling relations. Moreover, other studies such as \cite{baldi2018kinetic} or \cite{altamura2023galaxy} have focused on the kSZ effect, trying to disentangle the different components (namely the cluster peculiar velocity and rotation) producing the signal.

In view of the upcoming observational efforts, and as a continuation of the computational studies aforementioned, in this paper we aim to study the tSZ and kSZ effects on a sample of simulated galaxy clusters ($N \sim 10^2$) at different redshifts and dynamical states to provide results to interpret and lead future observations. For each simulated cluster in our sample we produce maps of the tSZ and kSZ signals. With these results, we study the mean features of both SZ effects, and look for possible correlations with the dynamical state of the clusters in our sample. We present the individual results for several prototypical clusters, together with a statistical approach considering the whole sample. The redshift evolution is also considered. 


The structure of the manuscript is a follows. In Sect. \ref{s:methods}, we give the details of the simulation, the halo finder and the cluster sample, and we describe how the dynamical state of the clusters is defined. In Sect. \ref{s:results}, we describe the results obtained for the analysis of single clusters and the whole sample.
Finally, we conclude and discuss our results in Sect. \ref{s:conclusions}.


\section{Methods}
\label{s:methods}

\subsection{The simulation}
\label{s:methods.simu}

In this work we examine a numerical simulation performed with the cosmological code \texttt{MASCLET} \citep{quilis2004masclet, Quilis_2020}. This code, designed primarily for cosmological purposes, employs (magneto-)hydrodynamics and $N$-Body techniques. The gaseous (collisional) component is evolved using high-resolution shock-capturing methods, while the DM component is evolved using an $N$-body multilevel Particle-Mesh (PM) scheme \citep{Hockney_Eastwood_1988}. The gravity solver ensures the coupled evolution of both components. Also, to enhance spatial and temporal resolution, \texttt{MASCLET} incorporates an adaptive mesh refinement (AMR) scheme.

The simulation encompasses a periodic, cubic domain with a size of $L=147.5 \; \mathrm{Mpc}$, discretised into $256^3$ cubical cells. It assumes a flat $\Lambda$CDM cosmology characterised by a matter density parameter $\Omega_m=0.31$ ($\Omega_\Lambda = 1 - \Omega_m$), a baryon density parameter $\Omega_b = 0.048$, and a Hubble parameter \mbox{$h \equiv H_0 / (100 \; \mathrm{km \; s^{-1} \; \mathrm{Mpc}^{-1}}) = 0.678$}. The initial conditions originate from a realisation of the primordial Gaussian random field, assuming a spectral index of $n_s = 0.96$ and an amplitude resulting in $\sigma_8 = 0.82$. These conditions are set up at redshift $z_\mathrm{ini}=100$ using a CDM transfer function \citep{Eisenstein_1998}. The chosen cosmological parameter values are in line with the latest findings reported by \cite{collaboration2020planck}.

Starting from the initial conditions and evolving them until the present time using a low-resolution run with equal-mass particles, we select regions that meet specific refinement criteria to establish three levels of refinement ($\ell = 1, 2$, and $3$) within the AMR scheme. In these initially refined levels, the DM component is sampled with particles that are 8, 64, and 512 times lighter than those employed for regions in $\ell = 0$. As the evolution progresses, local baryonic and DM densities are utilised to create finer grids, ultimately reaching a maximum refinement level of $\ell = 6$. The ratio between cell sizes for a given level ($\ell+1$) and its parent level ($\ell$) is $\Delta x_{\ell+1} / \Delta x_{\ell} = 1/2$, striking a balance between the prevention of numerical instabilities and the increase in resolution. This enables us to achieve a peak physical resolution of approximately $9 \; \mathrm{kpc}$ at $z = 0$. With four different DM particle species, the mass resolution amounts to approximately $2 \times 10^6 \; M_{\odot}$, equivalent to using $2048^3$ particles in the entire computational domain.

In addition to gravity, the simulation takes into account cooling mechanisms (free-free, inverse Compton, atomic and molecular cooling for primordial gas) and heating mechanisms (UV background radiation; \citealt{haardt1995radiative}).

\subsection{Cluster sample and dynamical state}
\label{s:methods.clusters}

\begin{figure}
\centering 
\includegraphics[width=1\linewidth]{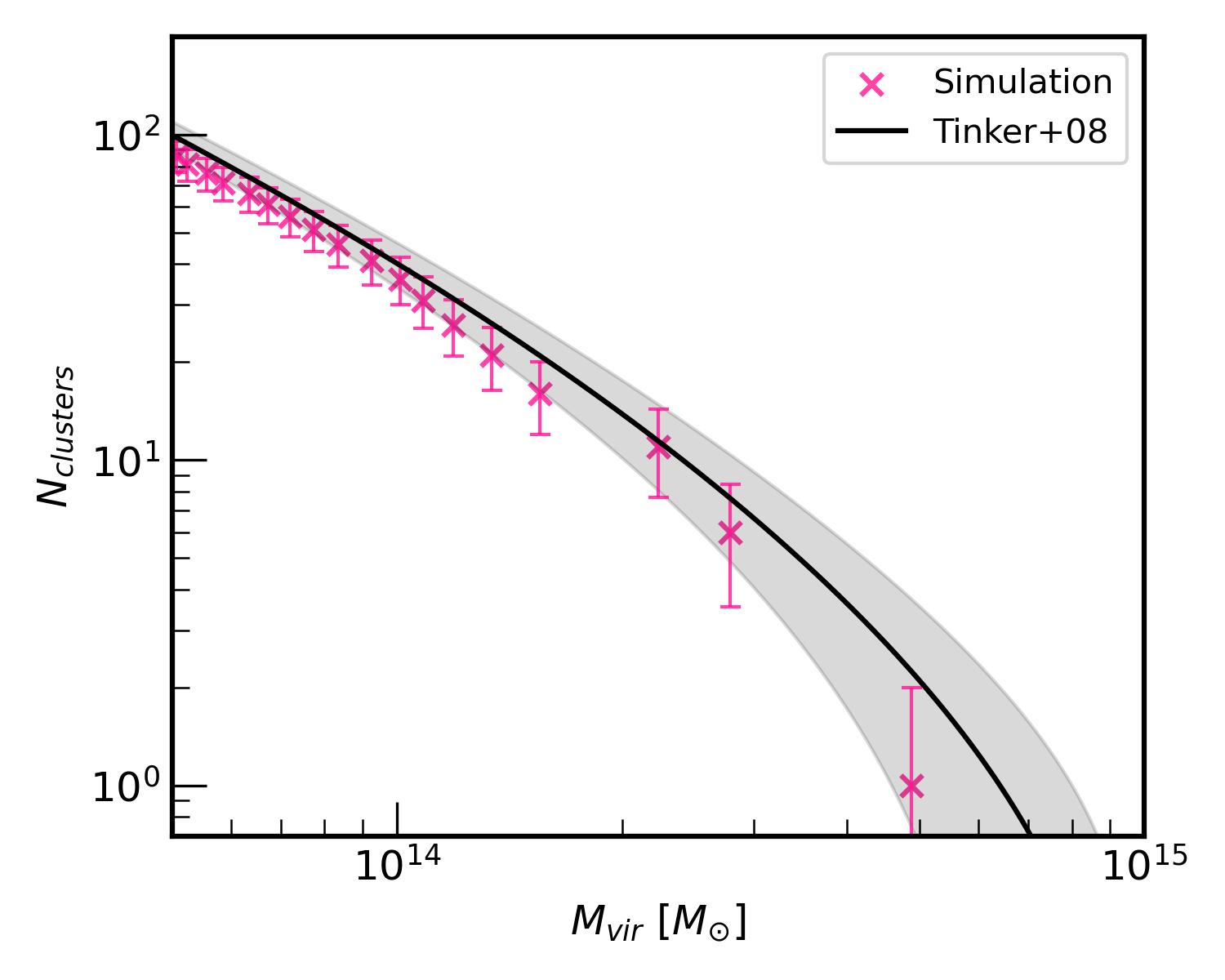}
\caption{Cumulative halo mass function at $z = 0$ of the simulation (points) and the fit (black line) described in \cite{tinker2008toward} for the cosmology given in Sect. \ref{s:methods.simu} using the \textsc{Colossus} package \citep{diemer2018colossus}. 
The errorbars and shaded region correspond to 1$\sigma$ Poissonian error.}
\label{fig:mass_function}
\end{figure}

In order to identify the GCs produced in the simulation, we use the publicly available spherical overdensity halo finder \texttt{ASOHF}\footnote{\url{https://github.com/dvallesp/ASOHF}} \citep[see][for more details]{planelles2010asohf, Knebe_2011, valles2022halo}. \texttt{ASOHF} finds DM bound structures (haloes) and substructures (subhaloes) in each snapshot of the simulation. In the simulation analysed in this work, \texttt{ASOHF} identifies $N = 86$ haloes with virial mass\footnote{The virial mass, $M_\mathrm{vir}$, is defined as the mass enclosed within a spherical region with radius equal to the virial radius, $R_\mathrm{vir}$, which is defined as the radius of a sphere containing an average overdensity of $\Delta_{\mathrm{c}} = \rho / \rho_c$. This critical overdensity for virialization, $\Delta_{\mathrm{c}}$, depends on redshift \citep{bryan1998statistical}. }, $M_\mathrm{vir}$, larger than $5 \times 10^{13} \; \mathrm{M_{\odot}}$ at $z = 0$. In Figure \ref{fig:mass_function}, we plot the halo mass function of this simulation in the last iteration against the fit provided by \cite{tinker2008toward}. In the $M_\mathrm{vir}/M_{\odot} \in [10^{13}, \; 5 \times 10^{14}]$ range fit and simulation are in a fair degree of agreement.

Since we want to study if and how the SZ effect changes depending on the dynamical state of the clusters, we need a robust estimation of this property. For this study, we employ the procedure described in \cite{valles2023choice}, in which a redshift-dependent combination of five different dynamical and morphological indicators is proposed to estimate the cluster dynamical state. While we refer the interested reader to the previous work, where a detailed description of the methodology employed to classify clusters according to their dynamical state is provided, here we briefly summarize the main procedure. We employ the following dynamical indicators of the DM halo:

\begin{itemize}
    \item Sparsity: $s_{200c,500c} = M_{200c} / M_{500c}$
    \item Ellipticity: $\epsilon = 1 - c/a$
    \item Centre offset: $\Delta_r = |\vec{r}_\mathrm{peak} - \vec{r}_\mathrm{CM}| / R_\mathrm{vir}$
    \item Mean radial velocity: $\langle \widetilde{v_r} \rangle_\mathrm{DM} \equiv \frac{\left|\langle v_r \rangle_\mathrm{DM}\right|}{V_\mathrm{circ,vir}}$
    \item Virial ratio: $\eta \equiv \frac{2E_\mathrm{k}}{|E_\mathrm{p}|}$
\end{itemize}
where $M_{200c}$ and $M_{500c}$ are the masses enclosed within the radius with average overdensity $200 \rho_\mathrm{c}$ and $500 \rho_\mathrm{c}$ respectively, $c$ and $a$ are the smallest and largest semi-axes of the best fitting ellipsoid describing the DM mass distribution, $\vec{r}_\mathrm{peak}$ and $\vec{r}_\mathrm{CM}$ are the position of the density peak and centre of mass with respect to the simulation box centre, $\left|\langle v_r \rangle_\mathrm{DM}\right|$ and $V_\mathrm{circ,vir}$ are the mean radial velocity of DM particles and the circular velocity at $R_\mathrm{vir}$, and $E_\mathrm{k}$, $E_\mathrm{p}$ are the total kinetic and potential energy within the same aperture.

Each indicator, $X_i$,  has a redshift dependent threshold, $X_i^\mathrm{thr}(z)$, below which the cluster is classified as relaxed. Therefore, for a particular cluster, if every indicator $X_i$ satisfies \mbox{$X_i < X_i^\mathrm{thr}(z)$} the cluster is regarded as \textit{totally relaxed}. Otherwise, the five dynamical indicators are combined to define a new parameter, $\chi$, representing the degree of \textit{relaxedness} of the considered halo \citep[see][for details on the definition of this parameter]{valles2023choice}. Following this procedure, a given cluster which does not fall into the \textit{totally relaxed} category will be classified as \textit{marginally relaxed} if $\chi\ge 1$ or \textit{unrelaxed} whenever $\chi< 1$. 

\begin{figure}
\centering 
\includegraphics[width=1\linewidth]{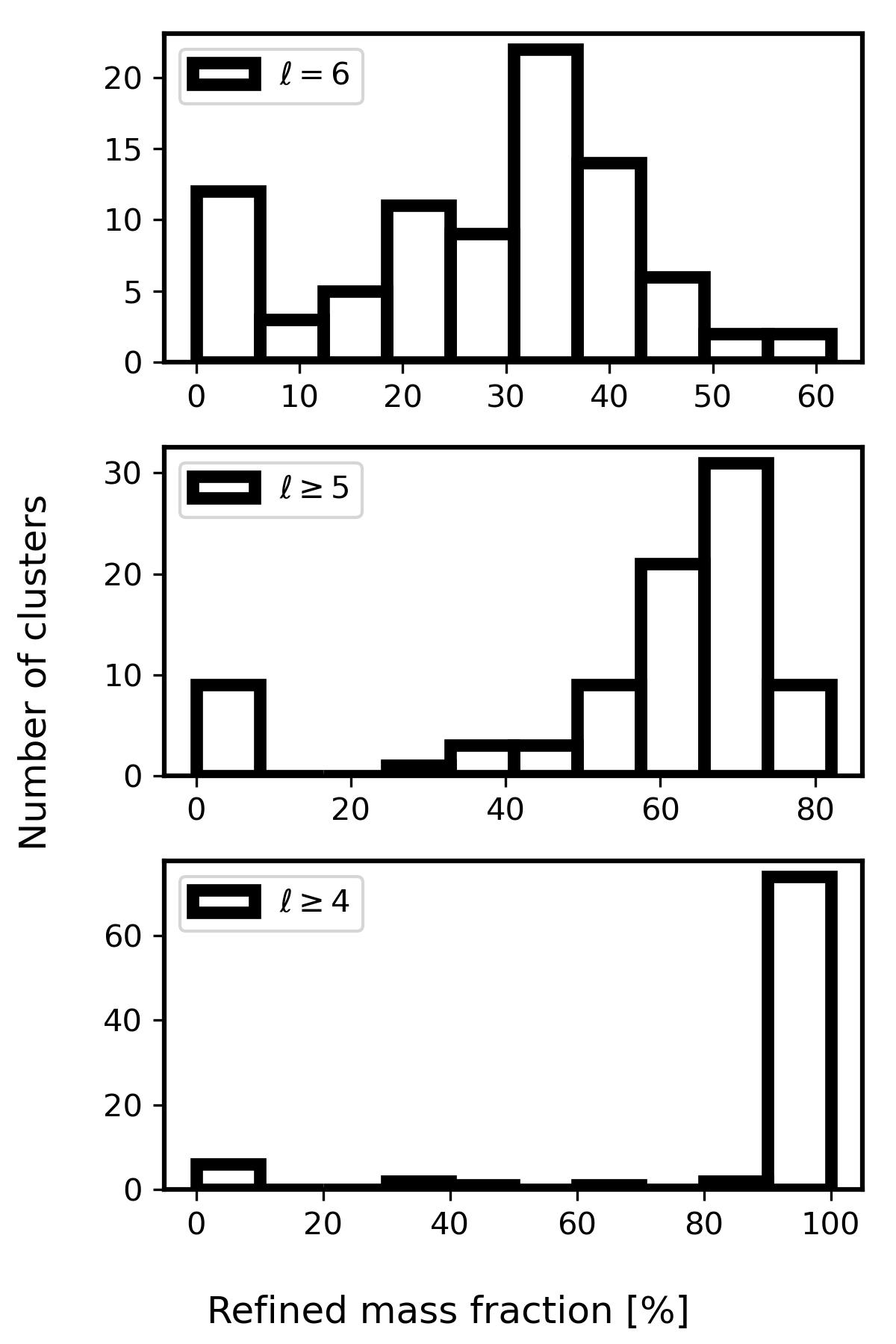}
\caption{Number of clusters in our sample as a function of the fraction of their mass, within $R_\mathrm{vir}$, that is refined at $\ell = 6$ (upper panel), $\ell \geq 5$ (middle panel) and $\ell \geq 4$ (lower panel). To avoid poorly resolved haloes from contaminating our results, in our final sample we have removed those haloes with $f_{\ell = 6} < 15 \%$, $f_{\ell\geq5} < 50 \%$ and $f_{\ell\geq5} < 90 \%$, $f_i$ being the corresponding mass fraction.}
\label{fig:mass_fraction}
\end{figure}

Applying this procedure to our cluster sample, we end up with $N_\mathrm{relaxed} = 11$, $N_\mathrm{unrelaxed} = 41$ and $N_\mathrm{marginally} = 34$ clusters at $z = 0$. However, in order to avoid contamination of our results due to poorly resolved haloes, for every cluster in our sample we have quantified the fraction of the cluster mass inside $R_\mathrm{vir}$ that is refined at $\ell = 6$, $\ell \geq 5$ and $\ell \geq 4$  (see Fig.~\ref{fig:mass_fraction}). According to these results, we have excluded from our sample all those haloes that fulfil that the mass fraction at each level, $f_i$, is: $f_{\ell= 6} < 15 \%$, $f_{\ell\geq5} < 50 \%$ or $f_{\ell\geq4} < 90 \%$. After pruning these poorly resolved haloes, we end up with a sample of  $N_\mathrm{relaxed} = 11$, $N_\mathrm{unrelaxed} = 20$, $N_\mathrm{marginally} = 32$ and, hence, a total of $N = 63$ clusters at $z = 0$, reducing our initial sample in a $25\%$.

On the other hand, we are also interested in studying how the SZ effect in a particular cluster evolves with $z$. For that purpose, we produce the merger tree from the \texttt{ASOHF} outputs and we trace the cluster population at $z = 0$ backwards in cosmic time, so that we can analyse the evolution of each of those $N = 63$ clusters individually.

\subsection{SZ mock $y$ maps}
\label{s:methods.SZ_calc}

There already exist packages to produce mock $y$ maps for the kinetic and thermal SZ effects, such as \texttt{pyMSZ} \citep{cui2018three, baldi2018kinetic}. Nevertheless, they are made for particle data in SPH simulations and, hence, we would need to properly transform our AMR data in order to use them. For this reason, we have designed our own package in order to produce mock SZ observations from block-based AMR simulations\footnote{\url{https://github.com/oscarmonllor99/MASC}} (in particular, for the \texttt{MASCLET} code). 

The algorithm works as follows. For each cluster in our sample, we define a map of physical size $2 R_\mathrm{vir} \times 2 R_\mathrm{vir}$ centred on the DM density peak, with the maximum resolution of the simulation. In the current version of our algorithm, the map line-of-sight has to be parallel to one of the three coordinate axes of the computational domain.
Once the map size is defined, we establish the integration width or depth, $w$, within which we integrate Eqs. \eqref{tSZ} and \eqref{kSZ}. In this case, we choose $w = 2 R_\mathrm{vir}$ and we centre the integration in the density peak so that the integration interval is $(- R_\mathrm{vir}, R_\mathrm{vir})$. Then, in order to calculate the $y$-value in each pixel, we integrate, using the rectangle rule, the former equations along the photon path with an integration step $\mathrm{d}s$ equal to the maximum resolution of our AMR grid. 
The integration is done at fixed $z$, that is, using just one snapshot for each case. This is because the photon crossing time, $\Delta t_{\gamma} \sim 1-10 \; \mathrm{Myr}$, is much smaller than the timescale for typical ICM motions, $\Delta t_{ICM}$, which we define as the timescale on which the ICM could change noticeably its structure at the considered scales due to its internal dynamics (see Fig. 3 of \citealt{valles2021troubled}). In this regard, since the gas moves at $v_\mathrm{ICM} \lesssim 10^3 \; \mathrm{km/s}$ \citep[e.g.][]{simionescu2019constraining} and $R_\mathrm{vir} \sim 1 \; \mathrm{Mpc}$, we have $\Delta t_\mathrm{ICM} \sim 1 \; \mathrm{Gyr}$. Thus, we should not expect the density, temperature and velocity fields to change significantly during the photon crossing time. 

Since our simulation does not take into account stellar feedback nor AGN, overcooling affects to small volumes of the clusters, leading to unphysical values of density and temperature, and therefore, of $y$ parameters.
To clear up this unwanted effect, we set up an upper comoving baryon density threshold for all cells involved in the integrated quantities along the line-of-sight. After some experimentation, we chose the value \mbox{$\rho_\mathrm{thres} = 5 \times 10^{-27} \; \mathrm{g\,cm^{-3}}$}. Finally, since the Compton scattering is significantly produced only in the hot-ICM phase, we eliminate from our calculation those cells with \mbox{$T < 10^{6} \; \mathrm{K}$}. With both thresholds we assure that all the cells contributing to the calculation have typical ICM values. In practice, for all clusters, only $< 2\%$ of the volume is dismissed.

To overcome the contained size of our sample, and only to increase the statistics when needed, we have 
considered the projections along the $x$, $y$ and $z$ axes, that is, projections on the $YZ$, $XZ$ and $XY$ planes, for each individual cluster as three different clusters in our extended sample. In this manner, the number of objects in the sample can be increased and get closer to the observational conditions in which no privileged direction is considered. 


\subsection{Mass normalization}
\label{s:mass_normalization}

In order to analyze the SZ signal dependence on the clusters' dynamical state, we should get rid of implicit dependencies on  other quantities like, for instance, mass. 
To do this, we need to find a mass dependence for each effect. For the thermal signal we have:
\begin{equation}
\label{tSZ_norm}
    y_\mathrm{tSZ} \propto \int \mathrm{d}s \; n_e T_e \propto \overline{\Sigma}_\mathrm{gas} \overline{T}_\mathrm{gas} \propto f(z) \frac{M_\mathrm{vir}^{5/3}}{R_\mathrm{vir}^2} \propto f(z) M_\mathrm{vir},
\end{equation}
where we have used that $\overline{\Sigma}_\mathrm{gas} \propto M_\mathrm{vir}/R_\mathrm{vir}^2$ is the mean surface density, the self-similar temperature-mass relation  \mbox{$\overline{T}_\mathrm{gas} \propto M_\mathrm{vir}^{2/3} f(z)$} for the mean ICM temperature \citep[e.g.][note that $f(z)$  is a function of redshift that is not relevant, since the normalization is done at fixed $z$]{voit2005tracing}  and the fact that, by definition, $R_\mathrm{vir} \propto M_\mathrm{vir}^{1/3}$. On the other hand, for the kinetic part we have:
\begin{equation}
  \label{kSZ_norm}
    y_\mathrm{kSZ} \propto \int \mathrm{d}s \; n_e v_\mathrm{los} \propto \overline{\Sigma}_\mathrm{gas} \overline{v}_\mathrm{los} \propto g(z) \frac{M_\mathrm{vir}^{3/2}}{R_\mathrm{vir}^{5/2}} \propto g(z) M_\mathrm{vir}^{2/3},
\end{equation}
where we consider that the average ICM velocity $v_\mathrm{los}$ scales as the circular velocity at the virial radius and, hence, \mbox{$\overline{v}_\mathrm{los} \propto g(z) \sqrt{G M_\mathrm{vir}/R_\mathrm{vir}}$}, being $g(z)$ a function of redshift, which will not contribute to our normalization, since it is done at fixed $z$.

Although based on simple arguments, the obtained scaling relations, Eqs. \eqref{tSZ_norm} and \eqref{kSZ_norm}, allow us to purge the mass dependence. As it is shown in the results presented in the next section, the clusters in our sample fit well to these scaling relations within the errors, except for the kSZ signal of relaxed clusters, that, as discussed in Sect. \ref{s:scaling_relations}, should not follow this mass dependence.


\section{Results}
\label{s:results}

In this section, we present the results obtained from the analysis of the mock SZ $y$ maps produced for every cluster in the sample. First, in Sect. \ref{s:results_z0}, we focus on the results at $z\sim 0$, and later, in Sect. \ref{s:results_evolution}, we extent the analysis to the effect of cosmic evolution on SZ effects. Throughout this work, unless otherwise specified, we use the term \textit{kinetic SZ effect} (kSZ) to refer to the effect produced when the monopole is subtracted, that is, the kinetic effect produced by the gas movements within the ICM in the cluster's reference rest frame.


\subsection{Analysis of the SZ effect at $z = 0$}
\label{s:results_z0}


Taking into account the dynamical state of the haloes in our sample, we analyse their associated SZ signals at $z=0$ by means of scaling relations (Sect. \ref{s:scaling_relations}), radial profiles (Sect. \ref{s:radial_profiles}) and projected maps (Sect. \ref{s:kSZ_maps}) of the most relevant quantities in order to find trends between different dynamical classes. We also try to disentangle the relevance of rotation on the kinetic effect (Sect. \ref{s:scaling_relations} and \ref{s:kSZ_maps}). Additionally, in Sect. \ref{s:projection_effect} we discuss how varying the line-of-sight direction can affect the signal (the projection effect), depending on cluster dynamical state. 

\subsubsection{Scaling relations}
\label{s:scaling_relations}

\begin{figure}[!h]
\centering 
\includegraphics[width=1\linewidth]{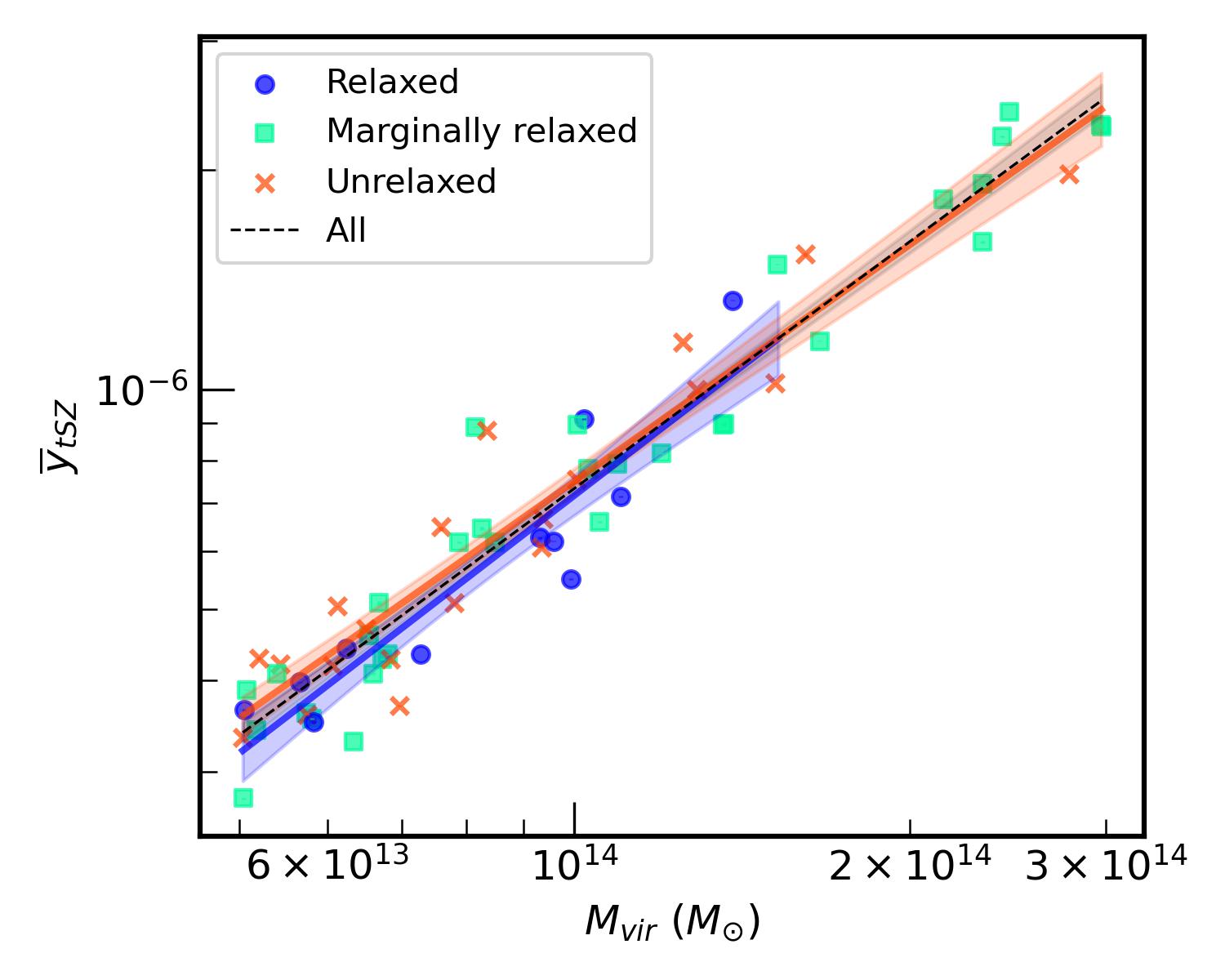}
\includegraphics[width=1\linewidth]{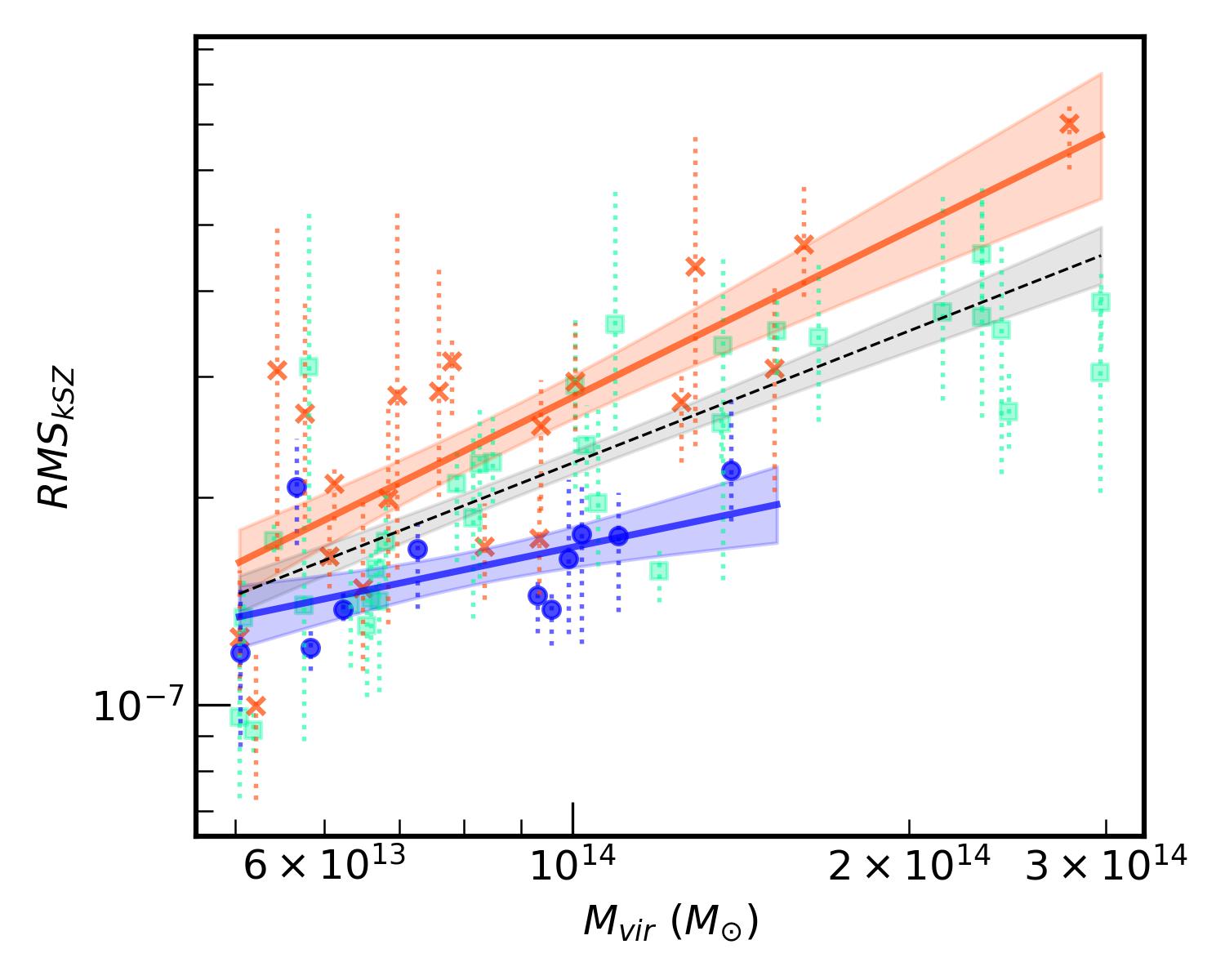}
\caption{Scaling relation between the virial mass, $M_\mathrm{vir}$, and the mean \mbox{$y$-parameter} for the tSZ effect (top panel) and for the RMS associated to the kSZ effect (bottom panel). The integration aperture is $R_\mathrm{vir}$. Represented points correspond, for each cluster, to the mean across the three spatial projections. In both panels, data have been fitted according to $\; y = A (M_\mathrm{vir}/M_0)^\alpha\;$, where $M_0 = 10^{14} \; M_{\odot}$ is the pivot mass and $A$ is a normalization.  The best fitting parameters are presented in Table \ref{tab:tab_fits}.  The shaded regions correspond to the $16-84$ confidence interval around the fit, whereas the error bars represent the variation (maximum and minimum values) across the three spatial projections for each cluster.}
\label{fig:scaling}
\end{figure}

\begin{table*}[!h]
    \centering
    \caption{Best fitting parameters obtained for Fig.~\ref{fig:scaling} using the fitting function $\; y = A (M_\mathrm{vir}/M_0)^\alpha$. The error corresponds to the standard error.}
    \begin{tabular}{|c|cc|cc|}
         \hline 
         & \multicolumn{2}{c|}{Thermal SZ} & \multicolumn{2}{c|}{Kinetic SZ} \\
         Dynamical state & $\alpha$ & $A$ & $\alpha$ & $A$ \\ \hline
         Relaxed & $1.16 \pm 0.17$ & $-22.44 \pm 2.45$ &  $0.35 \pm 0.15$ & $-8.71 \pm 2.81$ \\ 
         Unrelaxed & $1.08 \pm 0.09$ & $-21.24 \pm 1.22$ & $0.77  \pm  0.12$ & $-17.84 \pm 2.23$ \\ 
         All & $1.12 \pm 0.04$ & $-20.74 \pm 0.50$ & $0.63 \pm 0.07$ & $-15.14 \pm 1.10$ \\ 
         \hline
    \end{tabular}
    \label{tab:tab_fits}
\end{table*}

A well-known scaling relation is the one between the tSZ \mbox{$y$-parameter} \citep[e.g.][]{collaboration2013planck}, integrated within some specific radius (for example, $R_{500c}$), and the mass enclosed within that aperture. This relation has also been confirmed from a numerical point of view \citep[e.g.][]{planelles2017pressure}. In our case, we use this relation as a test to validate our results. 

One of the goals of the present paper is to look for similar correlations and scaling relations for the kSZ effect. However, the analysis of this effect requires some cautions to be taken. Given that the kSZ signal can have positive and negative values, mean values are not a well suited election for this analysis. So as to compare with tSZ, we define the root mean square (RMS) of the $y_\mathrm{kSZ}$ pixel values of the mock kSZ maps\footnote{Since the mean value of the kSZ signal tends to zero, the RMS is practically equal to the standard deviation and, hence, it has the physical meaning of quantifying how much the signal deviates from zero.}. In this manner, the kSZ is described with a positive quantity and is directly comparable to the mean of the $y_\mathrm{tSZ}$ pixel values of the tSZ map. Hence, we will use the mean $y$-parameter instead of the integrated quantity for the tSZ effect in order to be consistent\footnote{Note that the integral and the mean of the $y$-parameter inside some radius $R$ will differ just by the factor $1/(\pi R^2)$ corresponding to the integration area.}. 

The results are presented in Fig. \ref{fig:scaling}, where we show the scaling relation between $M_\mathrm{vir}$ and the mean $y_{\mathrm{tSZ}}$ (upper panel, denoted by $\overline{y}_{tSZ}$) and the RMS of the $y_\mathrm{kSZ}$ values (lower panel, denoted by $RMS_\mathrm{kSZ}$), both integrated in a $R_\mathrm{vir}$ aperture. The shown errobars correspond to the difference between the maximum and minimum values of $\overline{y}_{tSZ}$ and $RMS_\mathrm{kSZ}$ found in the three projections ($XY$, $XZ$, $YZ$) of each cluster. Furthermore, in order to capture the difference between the trends of different cluster dynamical states, we perform a fit of the type $\; y = A (M_\mathrm{vir}/M_0)^\alpha \;$ where $M_0 = 10^{14} \; M_{\odot}$ is the pivot mass. The best fitting parameters are presented in Table \ref{tab:tab_fits}. No mass normalization is carried out here.

As expected, we obtain a tight correlation between $\overline{y}_{tSZ}$ and $M_\mathrm{vir}$, without any significant difference due to dynamical state. The slopes are similar and consistent within the $16-84$ percentile errors shown, being also compatible with the theoretical $\alpha = 1$ in Eq.~\eqref{tSZ_norm}. Regarding the kSZ part, we observe more scatter, especially at lower mass, and we also obtain different trends depending on the cluster dynamical state, as shown by data fittings. The unrelaxed population shows a steeper ($\alpha = 0.77$) trend compared to the relaxed clusters ($\alpha = 0.35$), but, more importantly, it shows a higher $RMS_\mathrm{kSZ}$ value at all masses compared to the relaxed population. As expected, the marginally relaxed population is halfway of the other two. Note how the fits for unrelaxed and all clusters are compatible, within their errors, with the theoretical value $\alpha = 2/3$ given in Eq. \eqref{kSZ_norm}, while the relaxed fit is incompatible (within 1$\sigma$). Also note how the marginally relaxed population, which is the largest, follows almost perfectly the fit obtained for the whole sample (dashed line), which is the most consistent with the theoretical $\alpha = 2/3$.

\begin{figure}
\centering 
\includegraphics[width=1\linewidth]{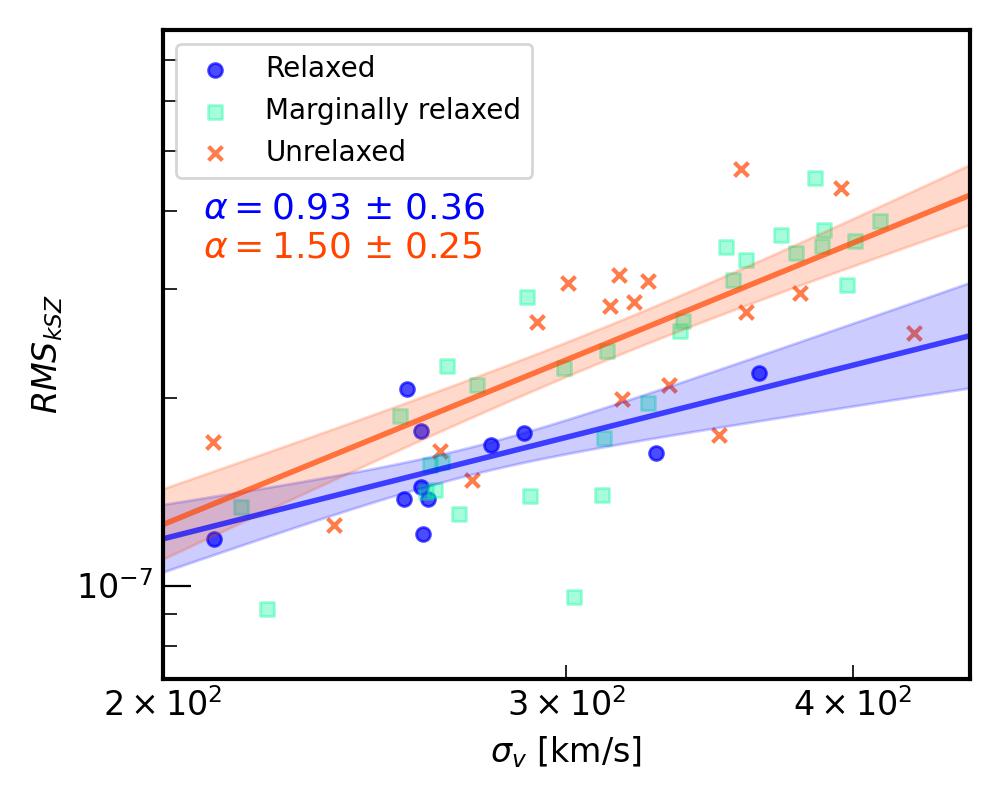}
\caption{ Scaling relation between the gas velocity dispersion inside the aperture and the root mean square for the kSZ. The integration aperture is $R_\mathrm{vir}$. The fit corresponds to $\; RMS_\mathrm{kSZ} = A (\sigma_v/\sigma_0)^B \;$ where $\sigma_0 = 300 \; \mathrm{km/s}$ is the pivot velocity dispersion and the shown error corresponds to a $16-84$ confidence interval. The fitting slopes ($\alpha$) and errors are shown below the figure legend.}
\label{fig:scaling_sigma}
\end{figure}

\begin{figure}
\centering 
\includegraphics[width=1\linewidth]{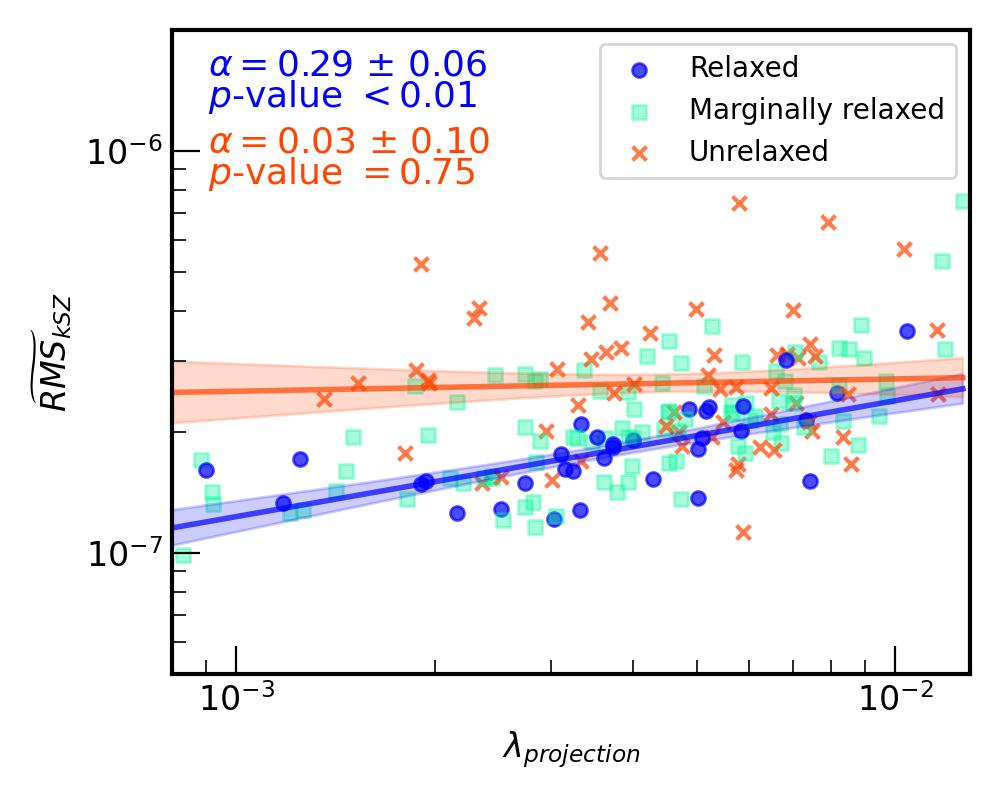}
\caption{ Scaling relation between the Bullock spin parameter ($\lambda_\mathrm{projection}$) of the angular momentum contained in each projection against the RMS of the kSZ signal of each projection normalized to a $M_\mathrm{vir} = 10^{14} \; M_{\odot}$ mass. Since neither of the two quantities depend on mass the relationship shown correlates the cluster spin with the kSZ effect. The fitting slopes ($\alpha$) and errors are shown in the upper left part.}
\label{fig:scaling_bullock}
\end{figure}

A plausible explanation for this separation in the kinetic effect, due to dynamical state, is the fact that the recent mergers and high accretion suffered by the unrelaxed clusters can trigger high velocity motions in the ICM (turbulence, outflows, etc.), and hence increase the overall kSZ signal of the cluster. In order to justify this fact, we show Fig. \ref{fig:scaling_sigma}, where we plot the kSZ signal against the gas velocity dispersion with an analogous fit. The higher the velocity dispersion is, the more intense the kinetic effect gets. On the other hand, the unrelaxed population is shifted towards higher velocity dispersion, while the relaxed sample is shifted towards lower velocity dispersion, hence having the unrelaxed population, on average, an overall higher kSZ signal. Moreover, at fixed $\sigma_v$, the value of $RMS_\mathrm{kSZ}$ for disturbed clusters is higher than for the relaxed ones. More precisely, we obtain, approximately, $RMS_\mathrm{kSZ} \propto \sigma_v^{3/2}$ and $RMS_\mathrm{kSZ} \propto \sigma_v$ from the disturbed and relaxed fittings, respectively.

As mentioned above, the slope of the fit of the relaxed population for the $RMS_\mathrm{kSZ}$ is incompatible within 1$\sigma$ error with the theoretical value ($\alpha = 2/3$). In order for a cluster of our sample to be relaxed, five different thresholds have to be fulfilled (see Sect. \ref{s:methods.clusters}). This means that our criteria for considering a cluster as relaxed is quite restrictive and, hence, their dynamics need to be really calm. In fact, relaxed clusters are mostly supported by thermal pressure (they are close to be in hydrostatic equilibrium) and the kinetic pressure produced by ICM bulk motions does not play an important role in its support against gravitational collapse. Thus, it is possible to expect that the scaling derived for the kSZ mass dependence in Eq. \eqref{kSZ_norm} is overestimating the value of $v_\mathrm{los}$, which for our relaxed population is likely to be mostly contributed by residual motions, causing the disagreement. In principle, this difference in the slopes between relaxed and disturbed clusters for the kSZ signal could be used to infer the cluster dynamical state depending on which of both trends is followed.

It is also interesting to study the connection between the kinetic effect and the cluster angular momentum. 
In order to mitigate the mass dependence for studying this correlation, we have to normalize the kinetic effect according to Eq. \eqref{kSZ_norm}\footnote{We define \mbox{$\widetilde{RMS}_\mathrm{kSZ} \equiv (10^{14}/M_\mathrm{vir})^{2/3} RMS_\mathrm{kSZ}$} as the RMS of the pixel values of the kinetic map normalized to \mbox{$M_\mathrm{vir} = 10^{14} \; M_{\odot}$}. Equivalently, according to Eq. \eqref{tSZ_norm}, we define  \mbox{$\widetilde{y}_\mathrm{tSZ} \equiv (10^{14}/M_\mathrm{vir}) \, \overline{y}_\mathrm{tSZ}$}}. Besides, it would be desirable to study an angular momentum related quantity which is independent of mass. Thus, we have used the spin parameter presented in \cite{bullock2001universal}:
\begin{equation}
\label{bullock}  
\hspace{3cm} \lambda = \frac{L}{\sqrt{2} M V R}
\end{equation}
where $L$ is the magnitude of the total angular momentum inside a given radius $R$, $M$ is the mass enclosed within that radius, and $V$ is the circular velocity at radius $R$. Since we are working with projections, we will try to correlate to the kinetic signal only the component of the total angular momentum that lies in the projection plane. For instance, in the $XY$ projection we will only consider $L = \sqrt{L_x^2 + L_y^2}$, since the $L_z$ component of the rotation will not contribute to the kinetic signal, as it does not have velocity component along the line-of-sight.

The results are presented in Fig. \ref{fig:scaling_bullock}, where we apply a similar fit to the data to disentangle the differences between populations. Note that, since we have three projections for each cluster, and the effect and angular momentum are different in each projection, we can treat each projection as an independent cluster signal, hence enhancing our statistics by a factor of three. Unrelaxed clusters hardly show any degree of correlation since the scatter is considerable and the fit is compatible with an horizontal line ($\alpha = 0.03 \pm 0.10$, with $p$-value $= 0.75$). On the other hand, for the relaxed case, we observe less scatter and a clear tendency to have a higher kSZ signal with increasing angular momentum ($\alpha = 0.29 \pm 0.06$), which is translated into a small $p$-value (that is, there exist correlation). Therefore, from this figure, we can infer that in the disturbed case, post-merger motions dominate over rotation and hence blur any possible correlation with angular momentum, while relaxed clusters are dominated by rotation, showing a clear correlation with $\lambda$. 

\subsubsection{Projection effects}
\label{s:projection_effect}

\begin{figure}
\centering 
\includegraphics[width=1\linewidth]{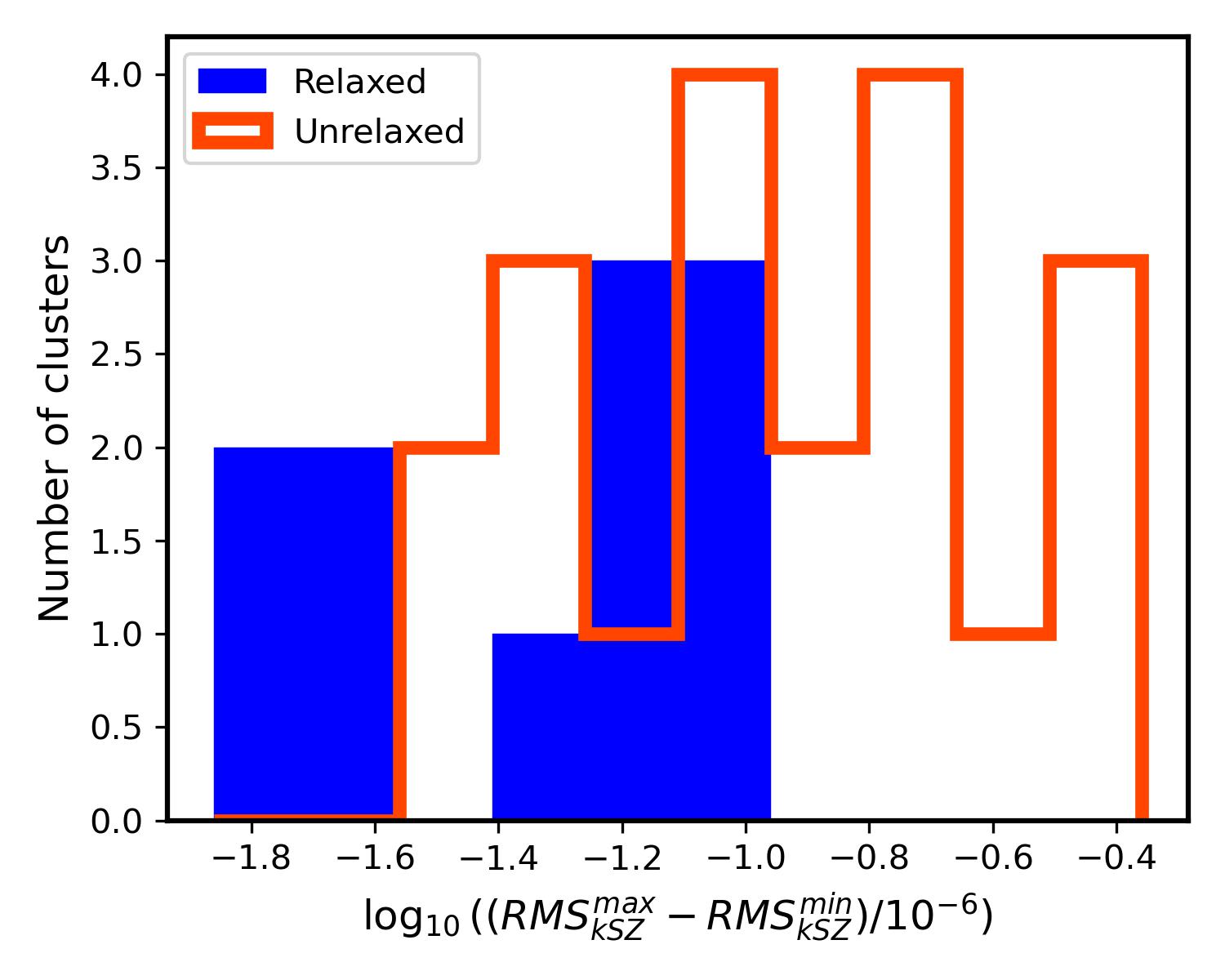}
\caption{Histogram of the number of clusters of each dynamical state class against the logarithmic difference between the maximum and minimum values of the RMS of the kSZ effect obtained with the three projections $XY$, $XZ$ and $YZ$.}
\label{fig:max_min_kinetic_difference}
\end{figure}

In Fig. \ref{fig:scaling}, we show for each cluster the average value of the $RMS_\mathrm{kSZ}$ when considering three different lines-of-sight aligned with the Cartesian axes of the computational domain, whereas the error bars show the range between the maximum and minimum values over the three different projections. It is noticeable  that the variation is in general significantly higher in disturbed clusters. In order to quantify this trend, in Fig. \ref{fig:max_min_kinetic_difference} we present an histogram of the variation of the RMS of the kSZ signal across the three different projections (lines-of-sight) for the relaxed and disturbed dynamical state classes. It is clear that low variations are dominated by relaxed clusters and large variations are mainly given in the unrelaxed cases. Hence, the impact of having the information projected in a certain plane is more important for disturbed clusters, since they tend to be aspherical and present significant substructure that can be erased due to projection effects. The opposite happens for relaxed clusters, as they generally have a rounder shape, minimizing the impact of projection.

For the tSZ signal, the projection effect is minimum compared to the kinetic part, to the point that the variation range is smaller than the data points in Fig. \ref{fig:scaling}. A plausible explanation comes from the fact that the kSZ signal depends on $v_\mathrm{los}$, which has a different value depending on the projection, while the tSZ effect involves scalar quantities which do not vary with the direction of observation.

\subsubsection{Radial profiles}
\label{s:radial_profiles}

\begin{figure}
\centering 
\includegraphics[width=1\linewidth]{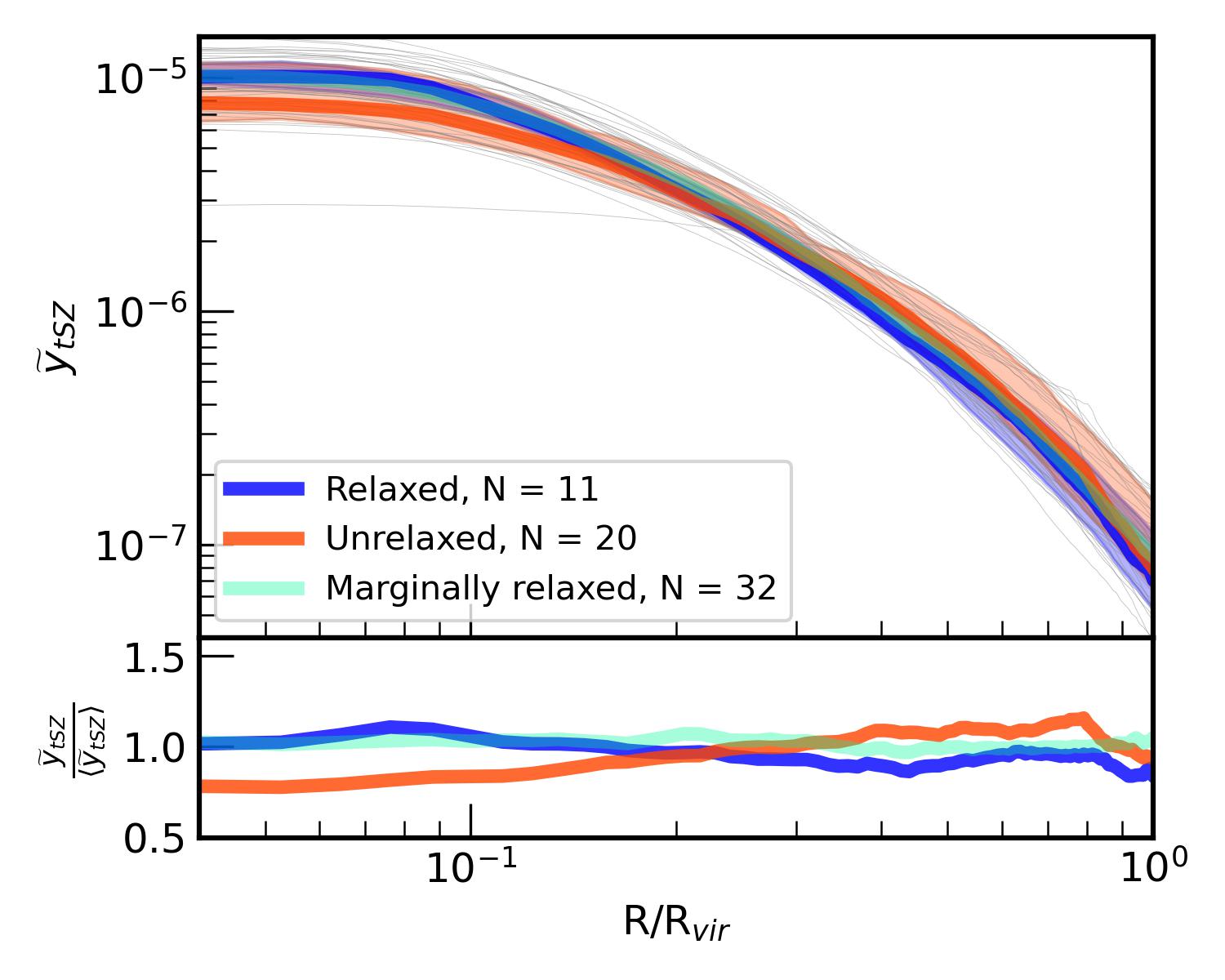}
\includegraphics[width=1\linewidth]{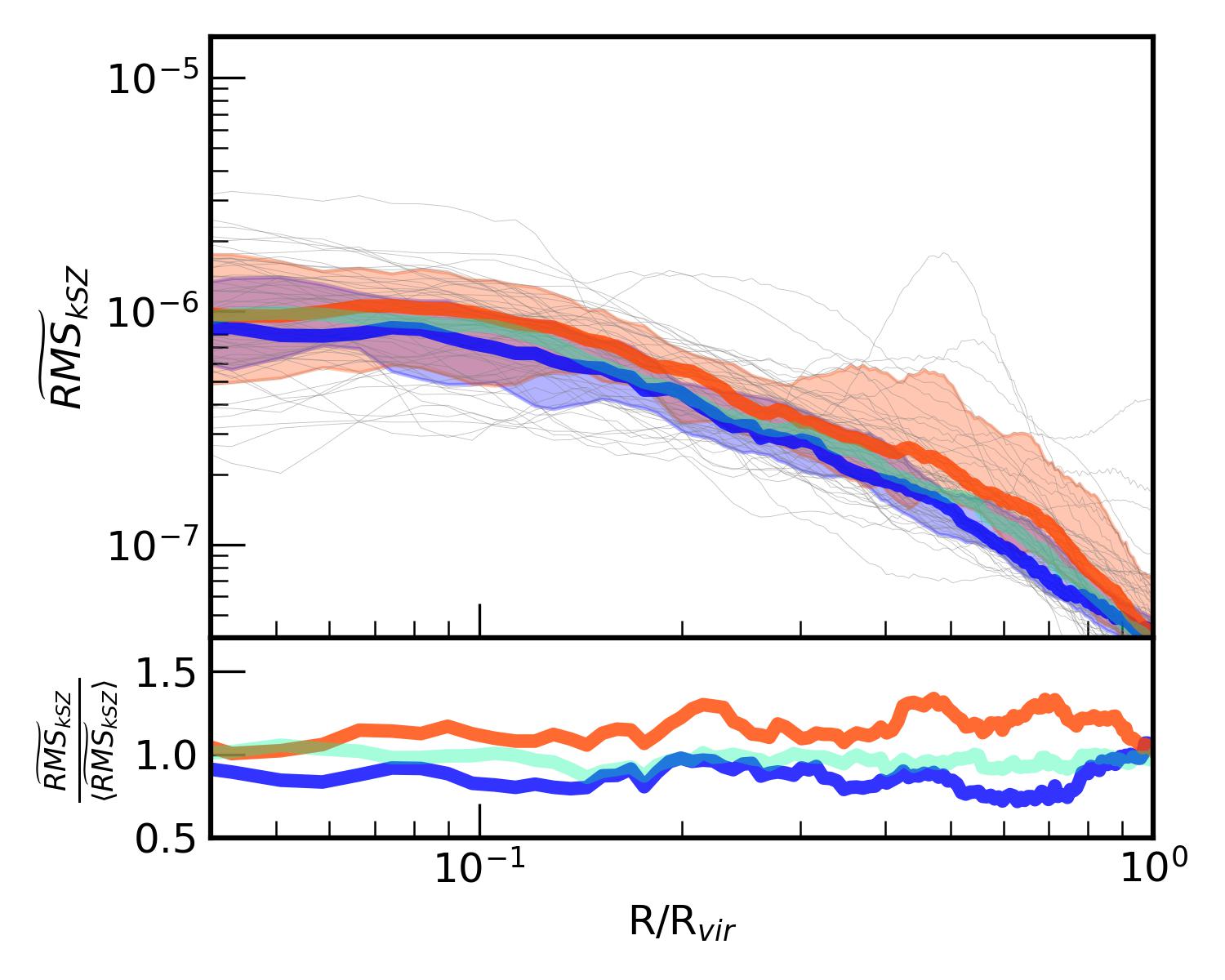}
\caption{Profiles at $z = 0$ of the $y$-parameter for the tSZ (top panel) and the kSZ (bottom panel) effects for different cluster populations. The thermal profile is obtained averaging over radial shells, whilst the kinetic profile is obtained computing its root mean square (RMS) in each shell. The solid lines represent the median of the profiles for each cluster population, whereas the shadowed areas correspond to a 68\% percentile. All clusters are normalized to the same mass, $M_\mathrm{vir} = 10^{14} M_{\odot}$.
Lower panels show the same quantity divided by the mean across the three dynamical state populations, thus highlighting the differences between them}
\label{fig:profiles}
\end{figure}

Another study to assess whether the cluster dynamical state affects the SZ signal would be to represent the radial profiles of this effect for each cluster population. In Fig. \ref{fig:profiles} we show the median profile at $z = 0$ of the $y$-parameter for the tSZ effect (top panel) and for the kSZ effect (bottom panel) for different cluster populations in terms of dynamical state. The thermal profile is obtained averaging over radial shells, whilst the kinetic profile is obtained computing its root mean square (RMS) in each shell. Since all clusters have been normalized to the same mass, $M_\mathrm{vir} = 10^{14} \; M_{\odot}$, we employ the symbols $\widetilde{y}$ and $\widetilde{RMS}$, respectively. This normalization ensures that the differences between populations are solely due to differences in the dynamical state. The shadowed regions represent the error corresponding to a 68\% percentile. The small panels below show the same quantity divided by the mean across the three dynamical state populations, thus highlighting the differences between them. 

For the tSZ effect we appreciate how the unrelaxed clusters have a flatter profile than the relaxed ones, which are more concentrated towards the cluster centre. Indeed, it can be seen that for $R > 0.2 \, R_\mathrm{vir}$ the unrelaxed clusters show a greater signal, whereas for $R < 0.2 \, R_\mathrm{vir}$ the relaxed clusters have a higher profile. This result is in agreement with recent observations carried out by \cite{adam2023xxl}, where they confirm that disturbed clusters have a relatively flatter core and a shallow outer slope of the pressure profile (note that $y_\mathrm{tSZ}$ is an indirect measure of the electron gas pressure). On the other hand, for the kSZ effect we observe that the unrelaxed population has a higher value for all radii, being the shape of the profile equivalent in all dynamical states. Marginally relaxed clusters show a behaviour that is halfway of the other two. The explanation for the described facts is similar to the one given in Sect. \ref{s:scaling_relations}: the thermal profile for unrelaxed clusters is less concentrated towards the centre due to, most importantly, recent merger(s), that move matter to the outer regions, hence rising there the SZ signal (both thermal and kinetic). Furthermore, turbulence and other motions caused by recent mergers suffered by the unrelaxed population produce high velocities that increase the overall kSZ signal, hence the overall higher profile for the unrelaxed population. Also, since marginally relaxed clusters are transitioning from unrelaxed to relaxed or vice versa, they show a middle term behaviour.

An important result given by theoretical studies \citep[e.g.][]{rephaeli1995comptonization} and observations \citep[e.g.][]{adam2017mapping} is the fact that the tSZ signal is approximately an order of magnitude larger than the kSZ signal, making the first one much easier to measure. This result is in agreement with the trend shown in Fig.~\ref{fig:profiles}, where the maximum values of the profiles satisfy that \mbox{$\widetilde{y}_{tSZ}\; / \; \widetilde{RMS}_{kSZ} \sim 10$}  at the innermost region. Nevertheless, we see that in cluster outskirts, at $R\geq R_\mathrm{vir}$, both effects become equally important, due to the higher importance of non-thermal motions in these regions.

\subsubsection{Kinetic SZ maps and multipole expansion}
\label{s:kSZ_maps}

\begin{figure*}[!h]
\centering  
\includegraphics[width=1\linewidth]{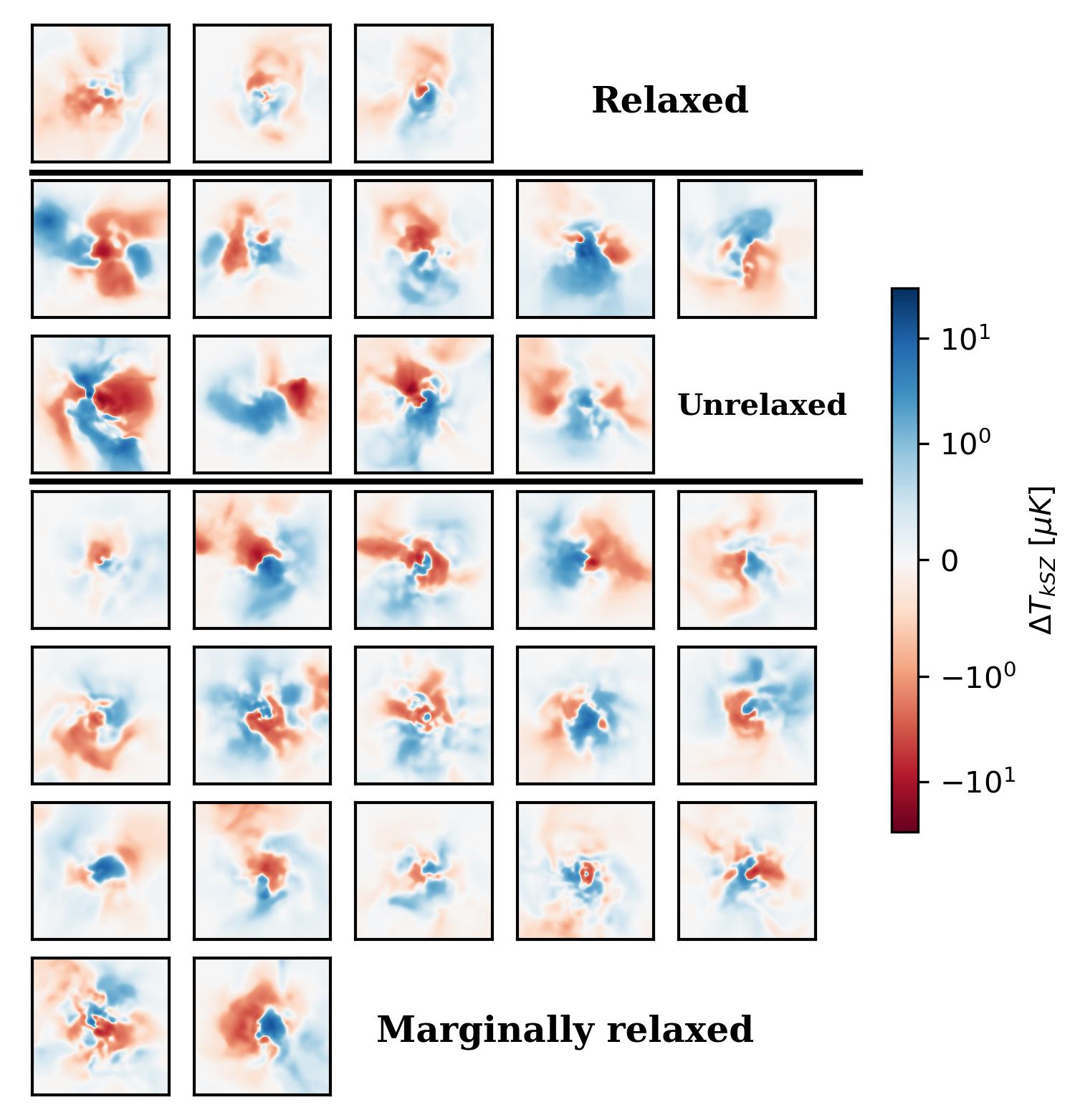}
\caption{$2R_\mathrm{vir} \times 2R_\mathrm{vir}$ maps of the kSZ effect for every cluster with $M_\mathrm{vir} > 10^{14} \; M_{\odot}$ at $z = 0$. The pixel size in the images is $9 \; \mathrm{kpc}$, coinciding with the highest numerical resolution in the simulation.  The maps are centred on the cluster density peak. On top there are the relaxed clusters, in the middle the unrelaxed and the marginally relaxed are left below. 
Colours display temperature variations according with the palette on the right.}
\label{fig:kSZ_maps}
\end{figure*}

In Fig. \ref{fig:kSZ_maps} we present the kSZ map for all clusters in our sample with \mbox{$M_\mathrm{vir} > 10^{14} M_{\odot}$} at $z = 0$, separated by dynamical state. The size of the maps is $2R_\mathrm{vir} \times 2R_\mathrm{vir}$ with the highest spatial resolution being $9 \; \mathrm{kpc}$ and they are centred on the density peak. It can be easily seen that the major difference between relaxed and unrelaxed maps is the fact that the first have the signal significantly concentrated towards the inner regions whilst the second have more features and present a more extended signal across the cluster volume. This explains, for example, the radial profiles shown in Fig. \ref{fig:profiles}, where the unrelaxed clusters clearly have a higher kSZ effect (and particularly towards the outer regions) or the scaling relation in Fig. \ref{fig:scaling} where the unrelaxed population show an overall higher kSZ signal for all masses. 

\begin{figure}[!h]
\centering 
\includegraphics[width=1\linewidth]{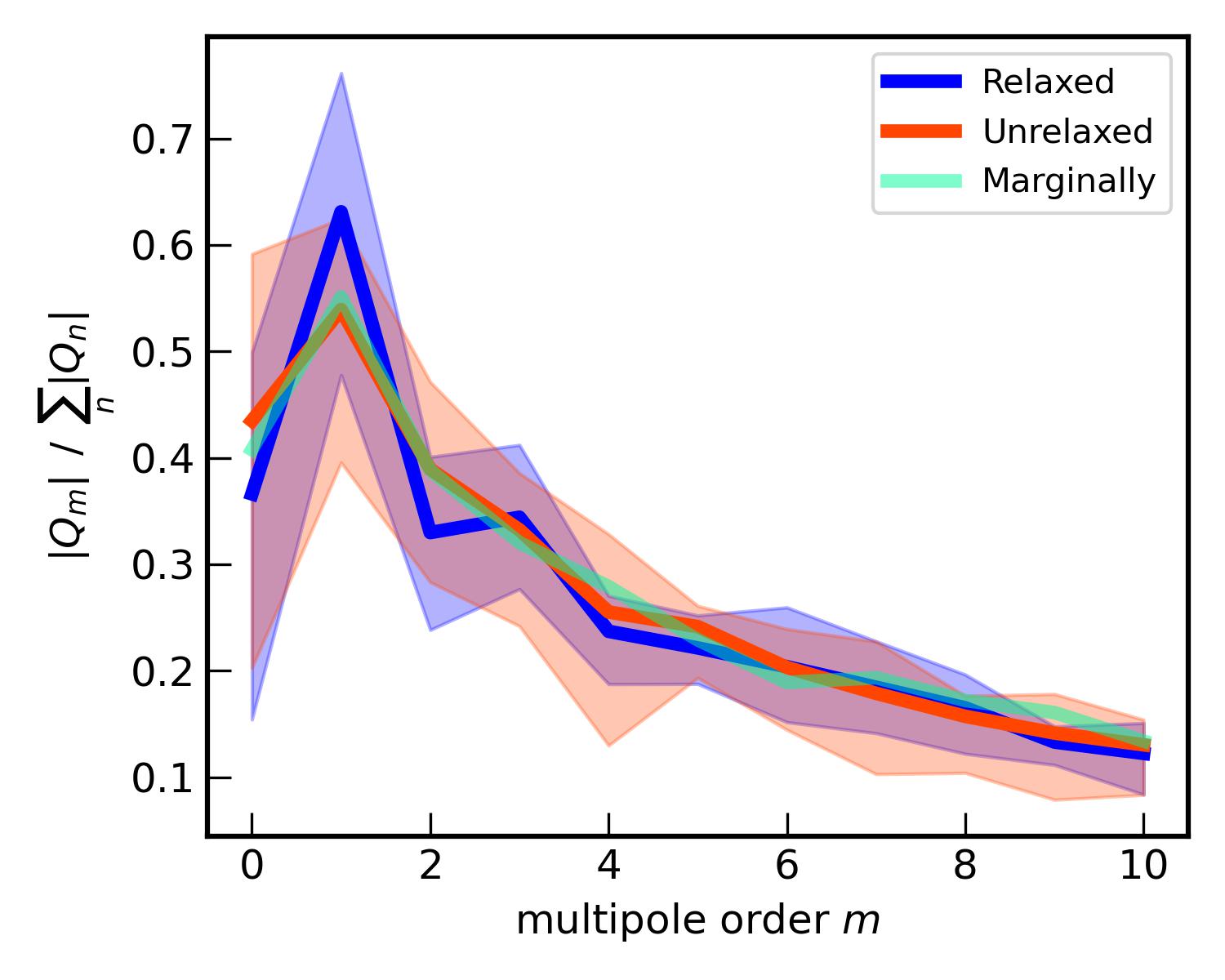}
\caption{Mean across dynamical state classes of the module of the multipole expansion coefficients of the kinetic SZ maps ($|Q_m|$) normalized by their sum against the coefficient order ($m$), for each projection. The integration aperture is $(0.1, \; 0.5) \, R_\mathrm{vir}$ and the error shows the 68\% confidence interval.}
\label{fig:rotation_importance}
\end{figure}

\begin{figure}
\centering 
\includegraphics[width=1\linewidth]{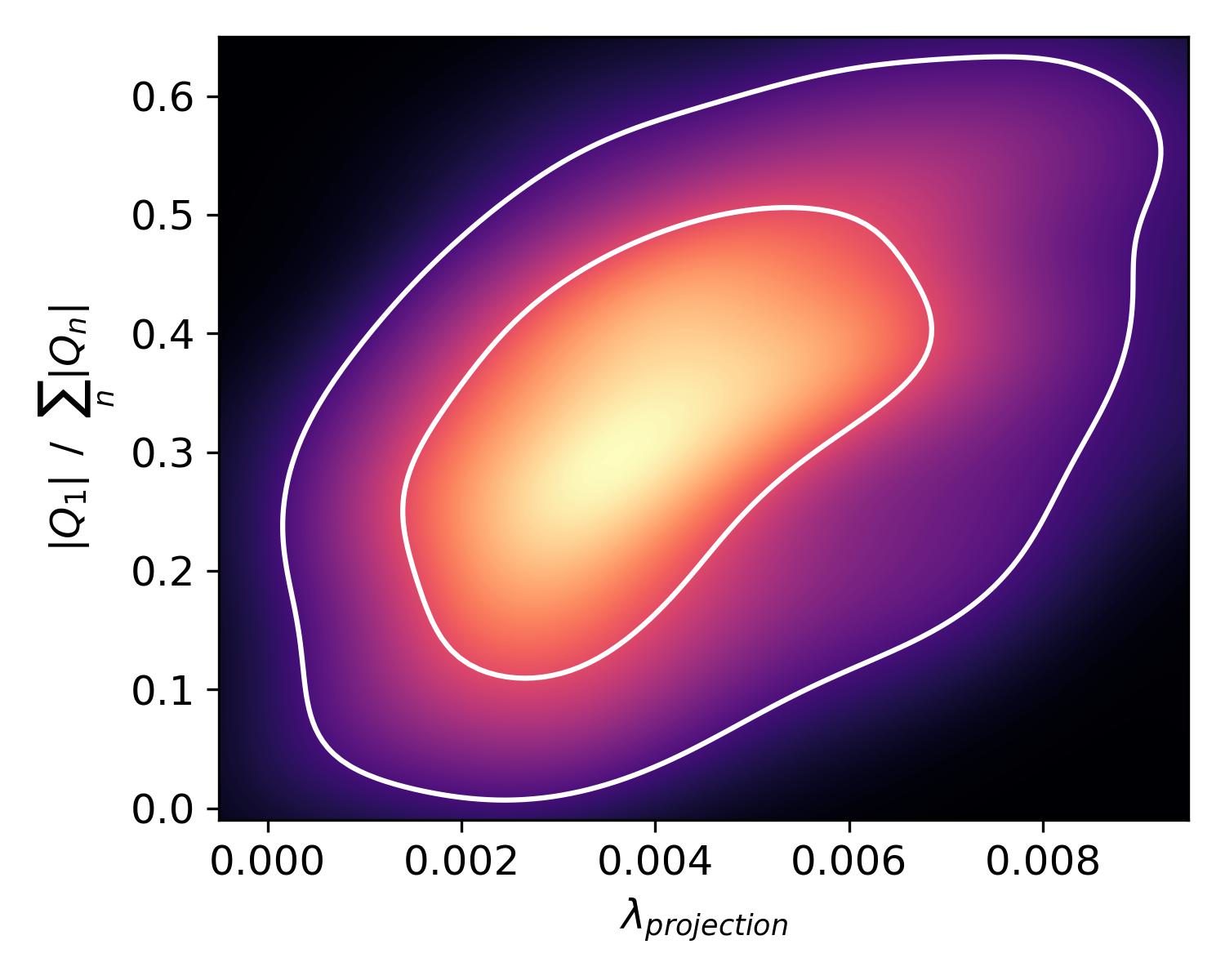}
\caption{Module of the dipole coefficient ($|Q_1|$) of the kinetic SZ map normalized by the sum of all the multipole expansion coefficients against the Bullock spin parameter ($\lambda_\mathrm{projection}$) of the angular momentum contained in each projection. The representation corresponds to a kernel density estimate (KDE) using all clusters in the sample.}
\label{fig:dipole_vs_rotation}
\end{figure}

\begin{figure*}[!h]
\centering 
\includegraphics[width=1\linewidth]{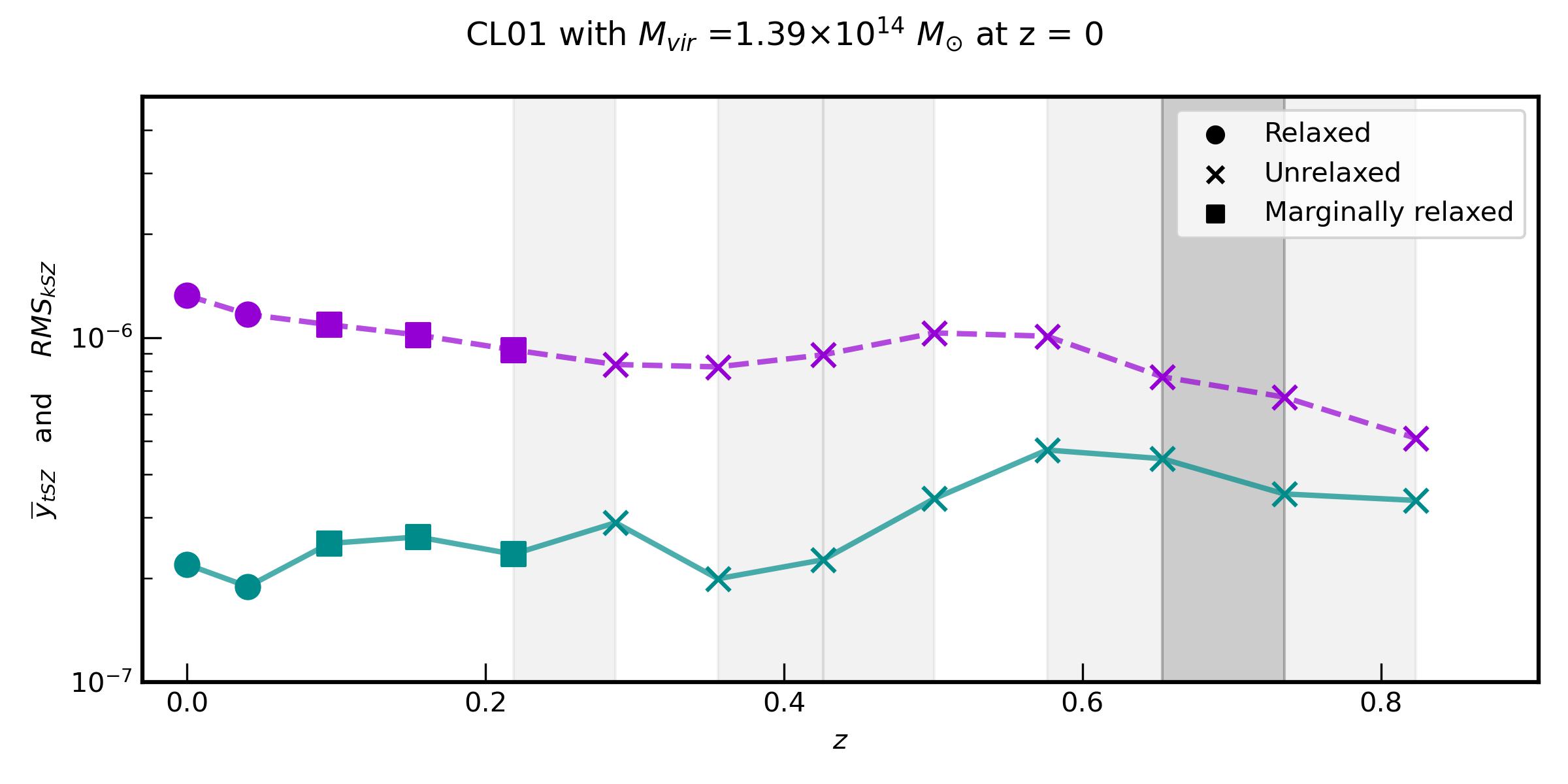}
\includegraphics[width=1\linewidth]{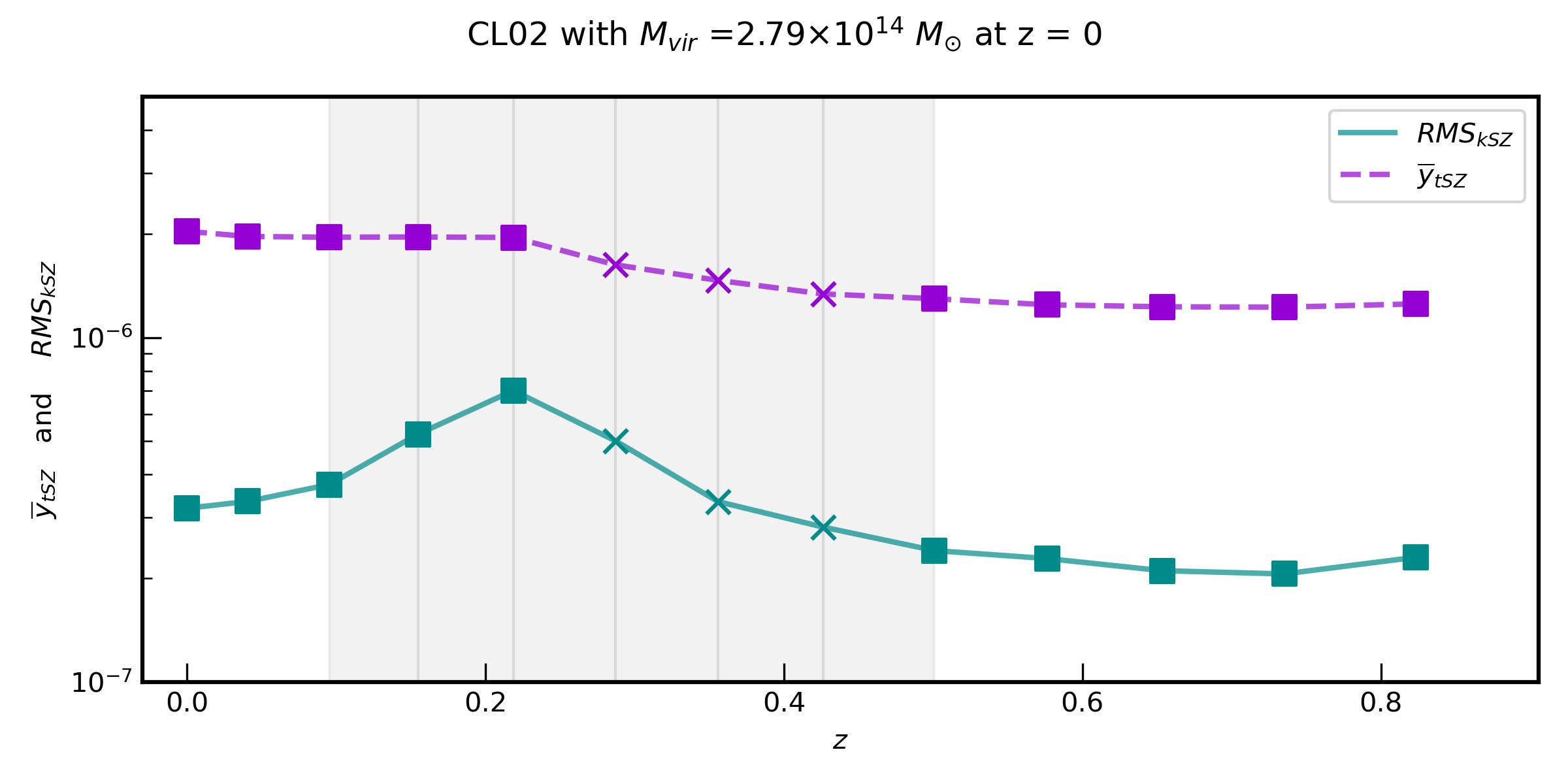}
\caption{Individual evolution of two prototypical clusters in the sample, CL01 and CL02 (top and bottom panel, respectively). In green (purple) we plot the integrated RMS of the kSZ signal (tSZ signal) as a function of redshift. Different symbols differentiate between the three stages of dynamical state and we also show in light (dark) grey the minor (major) mergers. 
Major (minor) mergers are highlighted as dark (light) grey regions.}
\label{fig:invidual_evolution}
\end{figure*}

On the other hand, because of cluster rotation, one should expect a dipole-like map once we have subtracted its bulk velocity and hence only internal motions remain \citep[e.g.][]{baldi2018kinetic, altamura2023galaxy}. Nevertheless, this is only appreciated in some maps, since the orientation of the angular momentum does not necessarily rest in the plane of the map and, in some cases, it could be even perpendicular, hence leaving the kSZ signal only to ICM turbulence, other bulk flows, etc. 

So as to quantify the importance of the rotation imprint in the kinetic SZ maps, we have performed a multipole expansion similar to that of \cite{gouin2020probing}, which is based on the methodology presented in \cite{schneider1997aperture}. In these studies the multipole expansion is applied to matter surface density, while in our particular case it is applied to the kSZ map. The goal is to obtain general angular symmetries for each dynamical state class. In particular, we are interested in discerning the importance of the rotation on the maps, whose imprint is a dipole-like signal \citep[e.g.][]{altamura2023galaxy}. The multipole moments of a certain quantity ($\Sigma$) defined on a polar map ($R, \phi$) are given by:

\begin{equation}
    \;\;\;\;\;\; Q_m (\Delta R) = \int_{\Delta R} R dR \int_0^{2\pi} d\phi e^{im\phi} \Sigma(R, \phi)
\label{eq:multipole}
\end{equation}
where $Q_m$ are the multipole moments and $\Delta R$ is the radial aperture over which we integrate. In our case, we have performed the integrals with $\Delta R = (0.1, \; 0.5) \, R_\mathrm{vir}$, where the dominant part of the effect is produced (see Fig. \ref{fig:kSZ_maps}).

The choice of the multipole expansion centre is crucial, as the centre of angular symmetries may not necessarily align with the image centre. Our choice for the new centre involves selecting the cell that corresponds to the arithmetic mean weighted by the absolute value of the signal. Through this straightforward recentring algorithm, we aim to minimise the introduction of a dipole signal in the expansion caused by the offset in the symmetry centre.

The results for the multipole expansion performed to all the sample kSZ maps are presented in \mbox{Fig. \ref{fig:rotation_importance}}, where we show the mean for each of the dynamical state classes of the module of the multipole expansion coefficients ($|Q_m|$) normalised by the sum of the complete multipolar expansion. On average, the trends of the three populations are similar, being the dipole coefficient ($m = 1$) clearly the most relevant, accounting for roughly $50 \%$ of the sum over all coefficients. This implies that, in general, the dipole (and, in principle, rotation) is the most important contribution to the effect. Nevertheless, the shown 68\% confidence intervals indicate that the weight of $Q_1$ highly varies across all clusters, due to projection effects. Moreover, as it can be seen, the $m = 2,3$ coefficients also contribute significantly to the signal and cannot be neglected. In fact, this is the main difference in the multipole expansion between disturbed and relaxed clusters: the first present a more prominent quadrupole behaviour and less dipole component. Considering the kSZ effect to be produced solely by rotation would result in a biased interpretation of the observations. 

\mbox{Fig. \ref{fig:rotation_importance}} shows an important feature of the kSZ signal that could be exploited by observations. Every kSZ map can show particular attributes but, when averaged over a significant number of clusters, the mean multipole expansion is dominated by the dipole and all the other coefficients become almost negligible, especially if only relaxed clusters are used. Hence, if properly recentred and oriented, many kSZ maps could be stacked and a dipole signal would emerge, which would be correlated to the global rotation properties of the sample. A similar procedure to the one described is applied in \cite{xia2021intergalactic} for intergalactic filaments.

For a more in-depth exploration of the rotational influence on the kinetic signal, it is crucial to know the extent to which the dipole is attributable to rotation versus other contributing factors. In Fig. \ref{fig:dipole_vs_rotation} we present a scaling relation for the whole sample between the module of the dipole coefficient ($|Q_1|$) of the kinetic SZ map (normalised by the sum of all the multipole expansion coefficients) against the Bullock spin parameter ($\lambda$) of the angular momentum for the three considered projections along the Cartesian axes. It can be seen that the dipole weight generally increases with increasing angular momentum in the corresponding projection. However, the scatter is considerable and, hence, apart from rotation, there are other ICM movements contributing to the dipole.

\subsection{SZ signal evolution with redshift}
\label{s:results_evolution}

In the present Section,  we study how the kinetic and thermal SZ effects evolve with cosmic time, looking for a correlation between the SZ signal variations and the clusters' dynamical state. First, we discuss the redshift evolution of the effect. Then, we analyse two particular cluster histories, in order to have a deeper description of this connection.

\subsubsection{Redshift evolution}
\label{s:z_evolution}
The SZ effect is essentially redshift independent (e.g. \citealt{carlstrom2002cosmology, mroczkowski2019astrophysics}). Nevertheless, our clusters evolve with redshift due to the expansion of the Universe and, hence, this introduces an \textit{artificial} redshift dependence that we have to take into account (e.g. see \citealt{schaan2021atacama} for similar arguments). We define the haloes of our simulations according to the over-density $\Delta = \frac{\rho}{\rho_B}$, where $\rho$ is the matter density and $\rho_B = \Omega_m \rho_c$ is the background density, with $\rho_c$ being the critical density of the Universe. Thus, since $\Delta$ is co-moving (independent of $z$), and $\rho_B = \rho_{B,0} (1+z)^3$, necessarily $\rho \propto \Delta (1+z)^3$.

On the other hand, the physical size of the cluster is bigger at lower redshift due to the Universe expansion. If $\mathrm{d}s$ and $\mathrm{d}x$ are the physical and co-moving distance elements, respectively, then $\mathrm{d}s = a(z) \mathrm{d}x$, where $a(z) = a_0/(1+z)$ is the scale factor at a given $z$. All in all, we have that the redshift evolution of the thermal effect is given by:
\begin{equation}
    \label{eq:z_evolution_tsz}
    y_\mathrm{tSZ} \; \propto \; \int n_e T_e \, \mathrm{d}s \; \propto \; \int \rho T_e \frac{\mathrm{d}x}{1+z} \; \propto \; (1+z)^2 \int  T_e \Delta  \, \mathrm{d}x 
\end{equation}
where $x$ is the co-moving line-of-sight distance and, thus, in the last step all the quantities inside the integral are comoving and, hence, redshift independent.


Similarly, the same procedure can be applied to the kinetic effect:
\begin{equation}
    \label{eq:z_evolution_ksz}
    y_\mathrm{kSZ} \; \propto \; (1+z)^2 \int v_\mathrm{los} \Delta \, \mathrm{d}x 
\end{equation}

We want to emphasise that this redshift dependence is introduced in our calculations only because of the manner in which we define these objects. That is, the factor $(1+z)^2$ is only taking into account how our clusters are evolving with cosmic time (density and size) due to the expansion of the Universe. Nevertheless, the same object with the same physical size $R$ and physical density $n$ at $z = 0$ and at arbitrary $z$ would produce the same effect at both redshifts, since there is no explicit $z$ dependence in Eq. \eqref{tSZ}. 

Therefore, in order to analyse the redshift evolution of the effect due to dynamical state changes, we will remove the \mbox{$(1+z)^2$} factor due to the intrinsic evolution of the cluster with the Universe expansion.

\subsubsection{Individual cluster evolution}
\label{s:individual_evolution}

Although difficult to elucidate due to the complexity of the mass assembly histories of GCs (e.g. \citealt{valles2020accretion}), it is clear from the previous results that there is a connection between the cluster dynamical state and the SZ signal (especially the kinetic part). To make this even clearer, we have chosen two prototypical clusters from our population to analyse how their dynamical state and both tSZ and kSZ signals change with cosmic time, taking into account whether the clusters suffer minor or major mergers. The results are presented in Fig. \ref{fig:invidual_evolution}, where we depict the cluster dynamical state, the kSZ (green) and tSZ (purple) signals and whether it has suffered a minor (major) merger in light (dark) grey bars.

In the first case, CL01 suffers a major merger at $ z \approx 0.7$, which rises the kSZ and tSZ signals. Once the merger phase ends at $z \approx 0.2$ it becomes marginally relaxed and, in the end, relaxed, decreasing its kSZ signal but maintaining the tSZ nearly constant or even increasing it, due to the mass growth. In the second case, CL02 starts with a major merger at $z \approx 1.1$, being thus an unrelaxed cluster. Once this period ends, it becomes marginally relaxed, reducing the kSZ signal until $z \approx 0.5$, where it enters a minor merger period, becomes unrelaxed and increases both, the thermal and kinetic signals. In the end, when it becomes again marginally relaxed, the kSZ effect decreases but the tSZ remains constant. 

Both study cases reassert the correlation between the cluster dynamical state (including mergers) and the kinetic part of the SZ signal. During merging periods (unrelaxed), the kSZ signal increases, while in quiescent periods (relaxed), it decreases. Simultaneously, the tSZ signal rises during merging periods due to mass gain and heating by compression and merger shocks, but does not decrease in relaxed periods. Therefore, in line with Fig. \ref{fig:scaling}, we can anticipate that, for the same mass, unrelaxed clusters should generally exhibit a more pronounced kSZ signal than the relaxed population, while the tSZ effect is minimally affected by this classification.

We can also highlight the thin line that can exist between our marginally relaxed an unrelaxed population with these particular clusters. Although CL01 only presents a marginally relaxed dynamical state once it ends the merging periods, CL02 exhibits this state even at the peak of the kSZ signal at $z \approx 0.2$, from which it starts to decrease. Hence, despite the fact that the marginally relaxed state appears in quiescent periods and at the end of merger or high accretion epochs, clusters can be significantly disturbed in this state. That is one of the main reasons why we predominantly highlight the differences between relaxed and unrelaxed populations in all the previous results, putting the marginally relaxed systems in a second plane, since they are halfway of the other dynamical states and can blur the main differences between them.

\section{Summary and conclusions}
\label{s:conclusions}

The main goal of this work is to assess the correlation between GCs dynamical state and their associated SZ signals, both thermal and kinetic. To do so, a sample of cluster-like haloes extracted from a cosmological AMR simulation is used. The SZ effect strongly depends on density, temperature and velocity distributions of the baryonic component of such haloes. Therefore,  those events able to substantially alter these quantities and, thus, changing the cluster dynamical state (e.g. mergers, accretion, etc.), would be crucial to establish the previously mentioned correlation.

We have defined three main dynamical state classes, according to \cite{valles2023choice}, and have primarily focused on the clearly differentiate cases of relaxed and unrelaxed (or disturbed) clusters, in order to maximise the main differences. Hence, at fixed redshift, we obtain the following results:

\begin{itemize}
    \item The kinetic part of the effect shows a strong correlation with cluster dynamical state. Disturbed clusters present an overall higher kSZ signal for all masses than the relaxed population and a steeper trend with mass ($\alpha = 0.77$ for disturbed vs $\alpha = 0.35$ for relaxed systems). This would be mostly explained by the fact that the unrelaxed sample has an overall higher velocity dispersion ($\sigma_v$) and, hence, the kinetic pressure plays an important role in the cluster dynamics. On the other hand, the relaxed population would be mostly supported by thermal pressure and, therefore, uncorrelated to the kinetic signal.
    
    \item Correlation between gas angular momentum and the kSZ effect is only evident among relaxed clusters, as the signal in disturbed objects would be dominated by  turbulence and non-rotational motions. By means of a multipole expansion on the kSZ maps, it becomes apparent that, overall, the dipole constitutes the most significant component of the effect, particularly within the relaxed sample. This characteristic can be leveraged by stacking multiple, appropriately oriented, and centred kSZ maps, allowing the retrieval of a dipole-like signal that correlates with the global rotation properties of the cluster sample.

    \item The thermal component lacks a robust correlation with cluster dynamical state, primarily because it is unaffected by the baryonic velocity distribution. However, the radial profiles presented seem to indicate that unrelaxed clusters exhibit a more flattened trend, while the relaxed sample displays greater concentration towards the cluster centre. This has been confirmed by observations in \cite{adam2023xxl}.
\end{itemize}

Our haloes at $z = 0$ can be traced backwards in cosmic time using the \texttt{ASOHF} merger tree. Hence, for each simulation snapshot we have recovered the cluster dynamical state history, together with the minor and major mergers suffered. Joining this information with the thermal and kinetic SZ signals of every cluster at all cosmic epochs we can infer that, during merging periods, when the cluster is disturbed, the kinetic part of the SZ effect rises, while in quiescent periods it decreases. The tSZ signal, however, does not decrease once the disturbed phase is over and thus, it seems to be almost independent of the cluster dynamical state.

Based on our findings, the kSZ effect opens a new window of opportunity to assess the dynamical state of clusters, particularly in distinguishing clusters undergoing a quiescent evolution from those marked by a history of significant merging events. Consequently, our results suggest that upcoming observational data, focused on measuring the SZ effect in clusters, could serve as a valuable complementary approach to describe the evolution of galaxy clusters. Additionally, and especially in cases where clusters exhibit a relaxed state, our analysis suggests that the kSZ effect holds potential for estimating dynamical quantities such as the gas rotation.


\begin{acknowledgements}
     This work has been supported by the Agencia Estatal de Investigación Española (AEI; grant PID2022-138855NB-C33), by the Ministerio de Ciencia e Innovación (MCIN) within the Plan de Recuperación, Transformación y Resiliencia del Gobierno de España through the project ASFAE/2022/001, with funding from European Union NextGenerationEU (PRTR-C17.I1), and by the Generalitat Valenciana (grant PROMETEO CIPROM/2022/49).
     OM and DV acknowledge support from Universitat de València through Atracció de Talent fellowships. Simulations have been carried out using the supercomputer Lluís Vives at the Servei d'Informàtica of the Universitat de València.
\end{acknowledgements}

\bibliographystyle{aa}
\bibliography{aa}

\end{document}